\documentclass{egpubl}
\usepackage{eurovis2024}

\SpecialIssuePaper         

\CGFccby

\usepackage[T1]{fontenc}
\usepackage{dfadobe}  

\usepackage{xcolor}
\usepackage{enumitem}
\usepackage{fontawesome}
\usepackage{colortbl}
\usepackage{array}
\usepackage{float}

\newcolumntype{R}[1]{>{\raggedright\arraybackslash}p{#1}}
\newcommand{\pheading}[1]{\noindent\textbf{#1}}
 
\arrayrulecolor{gray}

\usepackage{tabularx}
\usepackage{tabu} 
\newcolumntype{P}[1]{>{\arraybackslash}p{#1}}

\usepackage{xurl}

\usepackage{cite}  
\BibtexOrBiblatex
\electronicVersion
\PrintedOrElectronic
\ifpdf \usepackage[pdftex]{graphicx} \pdfcompresslevel=9
\else \usepackage[dvips]{graphicx} \fi

\usepackage{egweblnk}

\title[From Delays to Densities]%
      {From Delays to Densities: Exploring Data Uncertainty through Speech, Text, and Visualization}

\author[C. Stokes, C. Sanker, B. Cogley, \& V. Setlur]
{\parbox{\textwidth}{\centering Chase Stokes$^{1,2}$\orcid{0000-0001-7644-9021},
        Chelsea Sanker$^{3}$\orcid{0000-0001-6106-9587},
        Bridget Cogley$^{4}$\orcid{0009-0002-0125-5171},
        and Vidya Setlur$^{1}$\orcid{0000-0003-3722-406X}
        }
        \\
{\parbox{\textwidth}{\centering 
    $^1$Tableau Research, Palo Alto, USA\\
    $^2$University of California, Berkeley, USA\\
    $^3$Stanford University, Palo Alto, USA\\
    $^4$Versalytix, USA
       }
}
}

\begin{document}


\maketitle
\begin{abstract}
  Understanding and communicating data uncertainty is crucial for making informed decisions in sectors like finance and healthcare. Previous work has explored how to express uncertainty in various modes. For example, uncertainty can be expressed visually with quantile dot plots or linguistically with hedge words and prosody. Our research aims to systematically explore how variations within each mode contribute to communicating uncertainty to the user; this allows us to better understand each mode's affordances and limitations. We completed an exploration of the uncertainty design space based on pilot studies and ran two crowdsourced experiments examining how speech, text, and visualization modes and variants within them impact decision-making with uncertain data. Visualization and text were most effective for rational decision-making, though text resulted in lower confidence. Speech garnered the highest trust despite sometimes leading to risky decisions. Results from these studies indicate meaningful trade-offs among modes of information and encourage exploration of multimodal data representations. \\


\end{abstract}  

\section{Introduction}
\label{section:intro}

In today's world of data-driven decision-making, effectively communicating the uncertainty inherent to the underlying information is important~\cite{draper:1995}. \textit{Data uncertainty} refers to the range of potential outcomes, variability within a dataset, or possible error in measurements or predictions \cite{Morgan1990UncertaintyAG}. 
While precise data may be ideal for making decisions, such data is uncommon in real-life decisions. Communicating uncertainty can allow for a better understanding of the true state of the data. However, recognizing and effectively communicating data uncertainty poses several challenges. 

A key challenge in data uncertainty is interpretation. While experts might understand statistical nuances like confidence intervals or p-values, a lay audience might misread these indicators and make incorrect conclusions~\cite{Spiegelhalter:20111}. This misinterpretation significantly affects decisions in vital areas like medicine, finance, and public policy~\cite{Morss2008CommunicatingUI}. 
Conveying data uncertainty also affects trust, at times enhancing it~\cite{SteijaertSchaapRiet2021, sacha:2016}. 
However, uncertain data might also lead the audience to view the information as unreliable; despite uncertainty being a normal part of data analysis \cite{CapurroEtAl2021}. 

Researchers have developed various methods to assist in understanding statistical uncertainty. Visual tools like error bars, confidence intervals, and density plots illustrate data variability, scope, and distribution~\cite{MacEachren:2005,padillabook:2021}. In written texts, uncertainty is indicated through the use of hedge words such as ``somewhat'' and ``possibly''~\cite{lakoff:1973, deHaan2001}. In spoken communication, additional features like pitch and speech rate can signal a speaker's uncertainty or hesitation~\cite{BrennanWilliams1995, SchererEtAl1973, SwertsKrahmer2005}. 

There are tradeoffs, however, in the effectiveness of each mode of communication in conveying uncertainty. Viewers may not possess sufficient graphical literacy to understand complex visualizations~\cite{MacEachren:2005}, and textual explanations may be necessary to clarify visual nuances. On the other hand, readers might not follow lengthy explanations, resulting in limited comprehension~\cite{schriver:1997}. Finally, the transient nature of speech limits the ability to revisit information compared to text or visualizations~\cite{Sperber1995-SPER}. 

Determining how speech, text, or visualizations present information requires a nuanced understanding of how readers interpret information in each mode. A careful exploration of each mode can help identify effective techniques for data communication. As multimedia representations beyond visualizations emerge, it is important to develop a clear understanding of how other modes of information compare to visual representations. 

\pheading{Contributions.} 
To this end, we explore representations of data uncertainty (speech, text, and visualization) along a spectrum of concreteness to fuzziness to help identify effective strategies and trade-offs in communicating data uncertainty for decision-making. 
Two pilot studies outline a detailed design space for communicating data uncertainty. 
We focus specifically on unimodal representations to first understand what each mode can uniquely offer, an essential step before we can understand the implications of multimodality.

We conduct a comparative analysis of the efficacy and effects of these modes and fuzziness within each one, using two experiments. 
While this analysis partially replicates the methods of previous work, it also allows for direct comparison across modalities, leading to new findings. Unlike prior work, we examine these modes under a unified framework and uniform experimental conditions. 

The results indicate a high degree of trust in speech representations despite relatively irrational decision-making and a low rating of confidence in decisions made with text representations despite relatively rational decision-making.

\section{Related Work}
\label{section:related_work}

Our work builds on research in communicating uncertainty in data visualization, text, and speech, as well as decision-making.

\subsection{Communicating Uncertainty in Data Visualization}
Developing techniques for data visualization to address data uncertainty is an important research problem, as the interpretability of uncertain data can significantly impact decision-making~\cite{johnson:2003,hullmanerrorprone:2016,PADILLA2021275} in domains such as climate and geospatial modeling~\cite{MacEachren:2005,Pang2001VisualizingUI, Koo2015GeovisualizationOA,nagel:2019,ding:2020,witt:2023}, medicine~\cite{RISTOVSKI201460,hoefer:2022}, and business intelligence applications~\cite{vosough:2017}. For example, employing point estimates as part of uncertainty communication has been found to improve decision-making in contexts such as weather and transit. 
Surveys of uncertainty visualization~\cite{Bonneau2014OverviewAS,padillabook:2021,skeels:2008} and tools~\cite{jena:2020} for exploring the various approaches underscore the extensive scope and impact of research.

Important research has explored information visualization techniques for communicating uncertainty. 
Thomson et al. ~\cite{Thomson2005ATF} present a typology of uncertainty, delineating kinds of uncertainty matched with space, time, and attribute components of data. These concepts from visual semiotics are applied to characterizing visual signification appropriate for representing different categories of uncertainty~\cite{maceachren2012visual}. 
Research findings generally indicate that enhancing visualizations and data together with their uncertainty information improves users' understanding of complex information.

Some visualization techniques can better support data uncertainty. G{\"o}rtler et al.~\cite{gortler:2017} introduce bubble treemaps that encode uncertainty using wave-like modifications and blur effects. Sane et al.~\cite{Sane2021VisualizationOU} extend univariate confidence isosurfaces to multivariate feature level sets and visualize regions with uncertainty in relation to the specific trait or feature. Ensemble datasets contain a collection of estimates for each simulation variable, and ensemble visualizations representing a sample of projections better support understanding of data uncertainty~\cite{liu:2016,padillaensemble:2017}. 


\subsection{Communicating Uncertainty in Text}
Communicating uncertainty in text draws from research in linguistics, human-computer interaction, and information sciences~\cite{toulmin_2003}. Much of this work examines words and phrases that signal uncertainty in text, often called \textit{hedges}. Hedges indicate degrees of uncertainty and concreteness, such as `sort of,' `perhaps,' `might,' and `could be'~\cite{lakoff:1973, deHaan2001, SzarvasEtAl2012, rubinbook}. These words can indicate various kinds of uncertainty, e.g., probabilities for future events, matters of opinion, and information that is open to multiple interpretations~\cite{SzarvasEtAl2012, deHaan2001}. 

Computational linguistics research has implemented techniques for identifying hedging patterns to determine their effectiveness in communicating a particular point of view to the reader~\cite{islam-etal-2020-lexicon,goodluck:2021}. Hedge words decrease the perceived soundness of an argument \cite{BlankenshipHoltgraves2005,SmithClark1993} and the credibility of the author~\cite{SparksAreniCox1998}. Accordingly, participants are more likely to be persuaded by an argument without hedges \cite{DurikEtAl2008}. However, this finding may interact with how much confidence might reasonably be expected; descriptions of medical research are rated as more trustworthy when they include hedges \cite{Jensen2008}. The presence of hedges is often determined by inherent data uncertainty, as occurs in predictions of probabilistic future events~\cite{SzarvasEtAl2012, deHaan2001}.

\subsection{Communicating Uncertainty in Speech}
\label{rw_uncertainspeech}
In addition to the same lexical and semantic factors that are often studied in written text, the acoustic characteristics of speech provide an additional dimension to communicating uncertainty.

Confidence is correlated with faster speech \cite{SchererEtAl1973, SmithClark1993, BrennanWilliams1995}, 
higher intensity \cite{SchererEtAl1973}, and fewer, shorter pauses \cite{SchererEtAl1973}. Confident utterances are also more likely to exhibit falling intonation, while uncertain utterances are more likely to exhibit rising intonation \cite{SmithClark1993, BrennanWilliams1995, SwertsKrahmer2005}. Some studies find a lower pitch with more confident utterances \cite{JiangPell2015}, while others find a higher pitch \cite{SchererEtAl1973}.
These acoustic cues can also influence listeners' \textit{perception} of a speaker's certainty. Perceived confidence increases with faster speech rate \cite{BrennanWilliams1995, PonBarrySchieber2011, GuyerEtAl2019, kirkland2022s}, higher intensity \cite{JiangPell2017}, lower pitch \cite{JiangPell2017, GuyerEtAl2019, kirkland2022s}, falling pitch \cite{BrennanWilliams1995, LasarcykWollermann2010, GuyerEtAl2019}, fewer and shorter pauses \cite{LasarcykWollermann2010, PonBarrySchieber2011, JiangPell2017}. Predictors of perceived confidence have the same effects in synthetic speech \cite{kirkland2022s}.

Most of the existing work on the relationship between certainty and acoustic characteristics of speech addresses the \textit{speaker's} confidence, e.g., based on instructions to speak confidently or doubtfully \cite{SchererEtAl1973, JiangPell2015} or when answering trivia questions that they feel more or less confident about \cite{SmithClark1993, BrennanWilliams1995}. The work in perception of these acoustic characteristics similarly asks listeners to evaluate how confident a speaker sounds~\cite{BrennanWilliams1995, PonBarrySchieber2011, JiangPell2017, kirkland2022s}, rather than evaluating certainty of the information itself. 

In contrast, our work examines how acoustic characteristics, such as changes in speech rate, pitch, and pauses, might influence listeners' understanding of the uncertainty in probabilistic data and the way that they make decisions based on that data.

\subsection{Decision-making and Reasoning with Uncertain Information}
Perceived uncertainty and reliability of incoming information can influence the decisions that people make based on that information~\cite{hullmanpursuit:2019}. For instance, people are also more likely to seek more information when they are less confident in their decision about it~\cite{DesenderBoldtYeung2018}. 

There is evidence that some uncertainty visualizations (i.e., dot plots, probability density functions) allow people to better aggregate across varying estimates~\cite{greis:2018}. Kim et al.~\cite{kim2019bayesian, kim2020bayesian} looked into Bayesian reasoning scenarios, examining how different visualization techniques affect users' interpretive accuracy. Other research found that inferential uncertainty led to more overestimation of effect sizes than visualizations that show outcome variability~\cite{hofman:2020}. When uncertainty visualizations emphasize arithmetic means, users overlook uncertainty information and misinterpret visual distance as a proxy for effect size~\cite{kalevisualreasoning:2021}.  

Research has studied users' decision-making processes for specific domains. Korporaal et al. investigated how data uncertainty in maps might influence the process of spatial decision-making~\cite{Korporaal2020EffectsOU}.
Viewing COVID-19 visualizations with rising trends increased participants' beliefs that they and others were at risk~\cite{padillacovid19:2022}. More salient visualizations of uncertainty may lead to decreased willingness to follow COVID-19  forecasts~\cite{leffrang:2021}. 

In linguistically signaled uncertainty, the degree of confidence expressed in text or spoken utterance also influences decision-making. People are more likely to be persuaded by an argument that is expressed with moderate confidence than one that is expressed with very low or high confidence \cite{LondonEtAl1971}. Statements with moderate confidence are also rated as being more credible \cite{CramerEtAl2009}.

Specific acoustic correlates of confidence have been found to predict how listeners handle information; arguments given at faster speech rates are perceived as more credible and are more likely to persuade listeners than arguments with slower speech rates \cite{MillerEtAl1976, MehrabianWilliams1969}. Higher intensity and more variation in intensity are also positive predictors of how persuasive an utterance will be \cite{VanZantBerger2020}. Lower pitch produces more positive attitudes towards the message being communicated \cite{ChattopadhyayEtAl2003}. 


\section{Research Goals}
\label{section:research_goals}

The complexity of uncertain data necessitates the careful presentation and communication of information. Different modes of data representation can present different \textit{kinds} of information with tradeoffs in their effectiveness for communicating uncertainty. For example, speech information contains signals beyond what text alone can provide - the pitch and duration of certain words can communicate nuances that text cannot. Visualizations provide more information than can be concisely represented in text or speech formats. 

Our work aims to guide the effective use of these modes and provide insight for future work in multimodal uncertainty communication as well as situations where visual representations may not be feasible or useful to a given audience. We are specifically interested in the following research questions:

\begin{itemize}
    \item \textbf{RQ1}: \textit{How does the mode of information presentation affect decision-making with uncertain data?}
    \item \textbf{RQ2}: \textit{How does decision-making change when indicators of uncertainty are intensified within each mode of communication?}
\end{itemize}

We explored the design space of uncertainty representation in each mode of information, determining factors that increase perceptions of uncertainty. We then conducted two crowdsourced experiments. The first experiment compared how three different modes (speech, text, and visualization) communicate uncertain information. The second experiment compared decisions within each mode when those indicators of uncertainty were heightened.

\section{Design Space of Visualization, Text, and Speech Stimuli}
\label{section:design_space}

As part of investigating differences \textit{between} modes of communication, we varied the stimuli \textit{within} each mode on a scale from concrete to fuzzy, adapted from Setlur \& Cogley \cite{setlur2022functional}. 

Vagueness or \textit{fuzziness} occurs when the boundaries of meaning are not precisely defined, leading to ambiguity and multiple possible interpretations. Fuzziness is often unavoidable when distinctions are gradient rather than clear-cut categories. \textit{Concreteness} refers to a specific, clear, and unambiguous concept. Concrete representation relies on definite and precise terms that clearly delineate their meaning, leaving little room for ambiguity. Both concepts play a crucial role in communication, with fuzziness allowing for flexibility and adaptability, while concreteness ensures clarity and precision \cite{Sorensen1997, setlur2022functional}. In the context of this paper, we use ``precision level'' to refer to the variable capturing the fuzziness of stimuli.

\begin{table*}[ht]
\renewcommand{\arraystretch}{1.25}
\caption{Stimulus categories and examples for Experiment 1 ($2 \times 3$ study design). Hedge terms in the fuzzy text template are bolded to highlight the differences here but did not receive any visual treatment within the experiment.}
\label{tab:e1_stimuli}
\begin{tabular}
{|l|R{0.4\linewidth}|R{0.46\linewidth}|}
\hline
& \textbf{Concrete} & \textbf{Fuzzy} \\ 
\hline
\textbf{Visual} & 
 \raisebox{-.85\totalheight}{\includegraphics[width=0.25\textwidth]{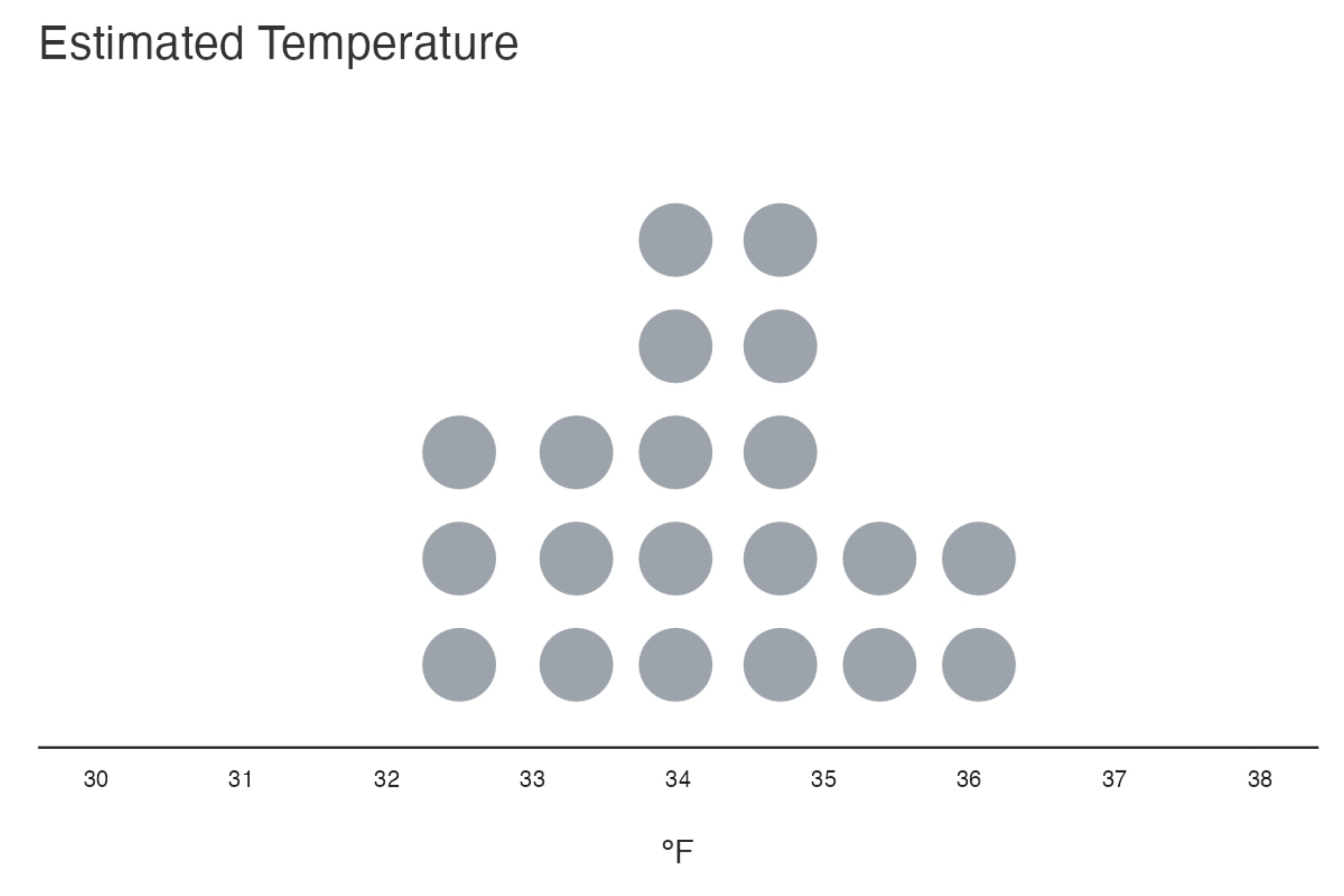}}
&  
 \raisebox{-.85\totalheight}{\includegraphics[width=0.25\textwidth]{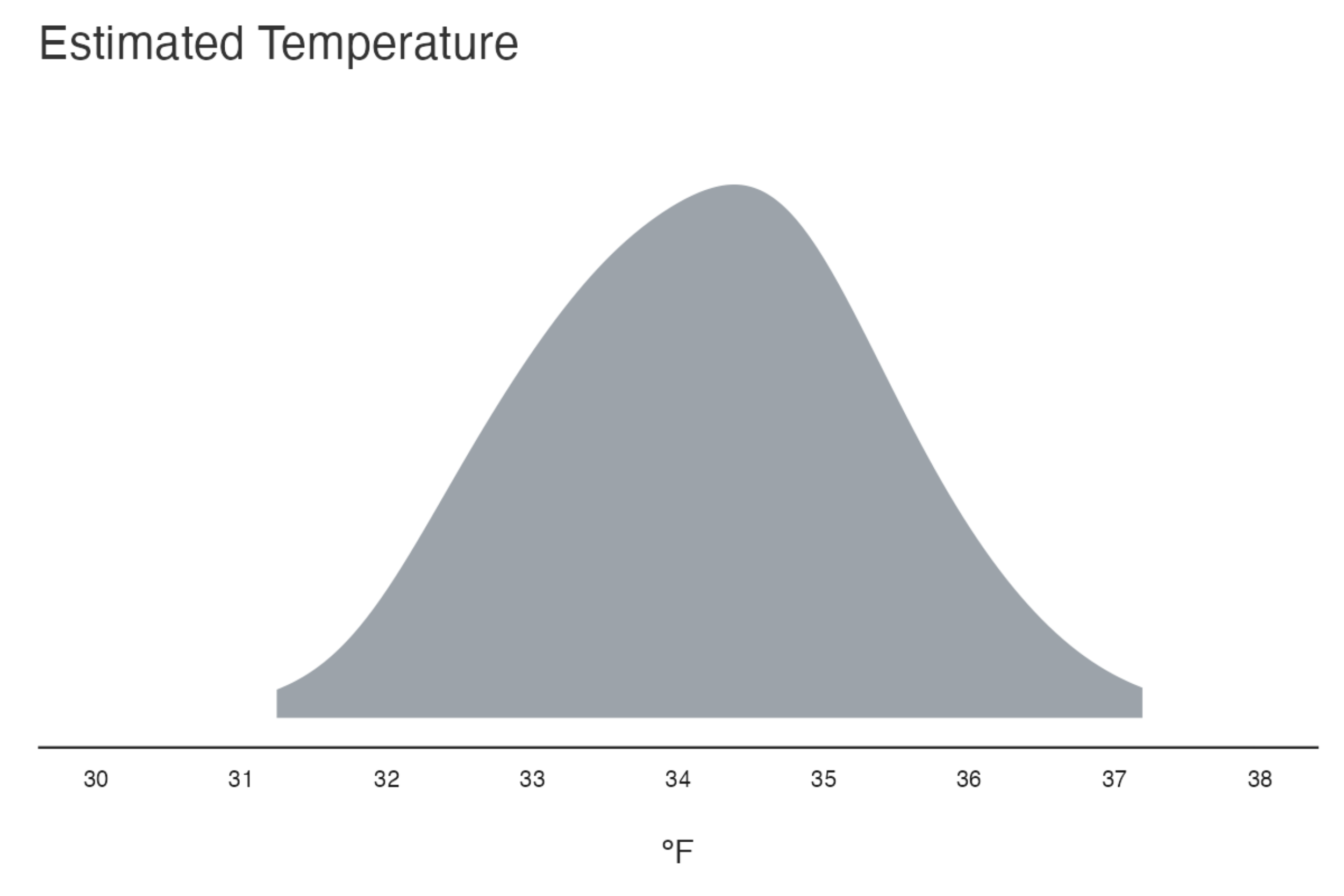}}
 \\ 
\hline
\textbf{Text} & 
The most likely temperature low tonight is [mean]ºF. There is a 50\% chance that the temperature will fall between [50\% interval min] and [50\% interval max]ºF. While the range of possible lows spans [min] to [max]ºF, those extremes are less likely. It is also [significantly/no modifier/somewhat/slightly] more likely to be [warmer than cooler/cooler than warmer] within that range. & 
The most likely temperature low tonight \textbf{might} be around [mean]ºF. There is a 50\% chance that the temperature \textbf{could} fall between [50\% interval min] and [50\% interval max]ºF. While the range of possible lows \textbf{could potentially} span [min] to [max]ºF, those extremes \textbf{seem} less likely. It also \textbf{appears} [significantly/no modifier/somewhat/slightly] more likely to be [warmer than cooler/cooler than warmer] within that range.  \\ 
\hline
\textbf{Speech} & Baseline Speech (\textcolor{blue}{\href{https://osf.io/3etjq?view\_only=78ba00b533b0485f9ba385e8f719693a}{MP3 Link}}) &  0.2s delay prior to numerical value, 65\% speed on numbers, 70\% speed on hedge or likelihood terms, 5\% pitch decrease on numbers and hedge or likelihood terms (\textcolor{blue}{\href{https://osf.io/gufh9?view\_only=78ba00b533b0485f9ba385e8f719693a}{MP3 Link}}) \\ 
\hline
\end{tabular}
\end{table*}

\begin{table*}[ht]
\renewcommand{\arraystretch}{1.25}
\caption{Stimulus categories and examples for Experiment 2. Cells with a light gray background indicate stimuli that overlapped with Experiment 1 (E1). A sample of the text stimuli is provided, and visual treatments shown here were applied to the stimuli in the survey. 
}
\label{tab:e2_stimuli}
\begin{tabular}
{|R{0.1\linewidth}|R{0.15\linewidth}|R{0.15\linewidth}|R{0.15\linewidth}|R{0.15\linewidth}|R{0.15\linewidth}|}
\hline
& \textbf{Most Concrete} & \textbf{Somewhat Concrete} & \textbf{Mixed} & \textbf{Somewhat Fuzzy} & \textbf{Most Fuzzy} \\ 
\hline
\textbf{Visual} 
& \cellcolor[HTML]{f2f2f2}\raisebox{-.85\totalheight}{\includegraphics[width=0.13\textwidth]{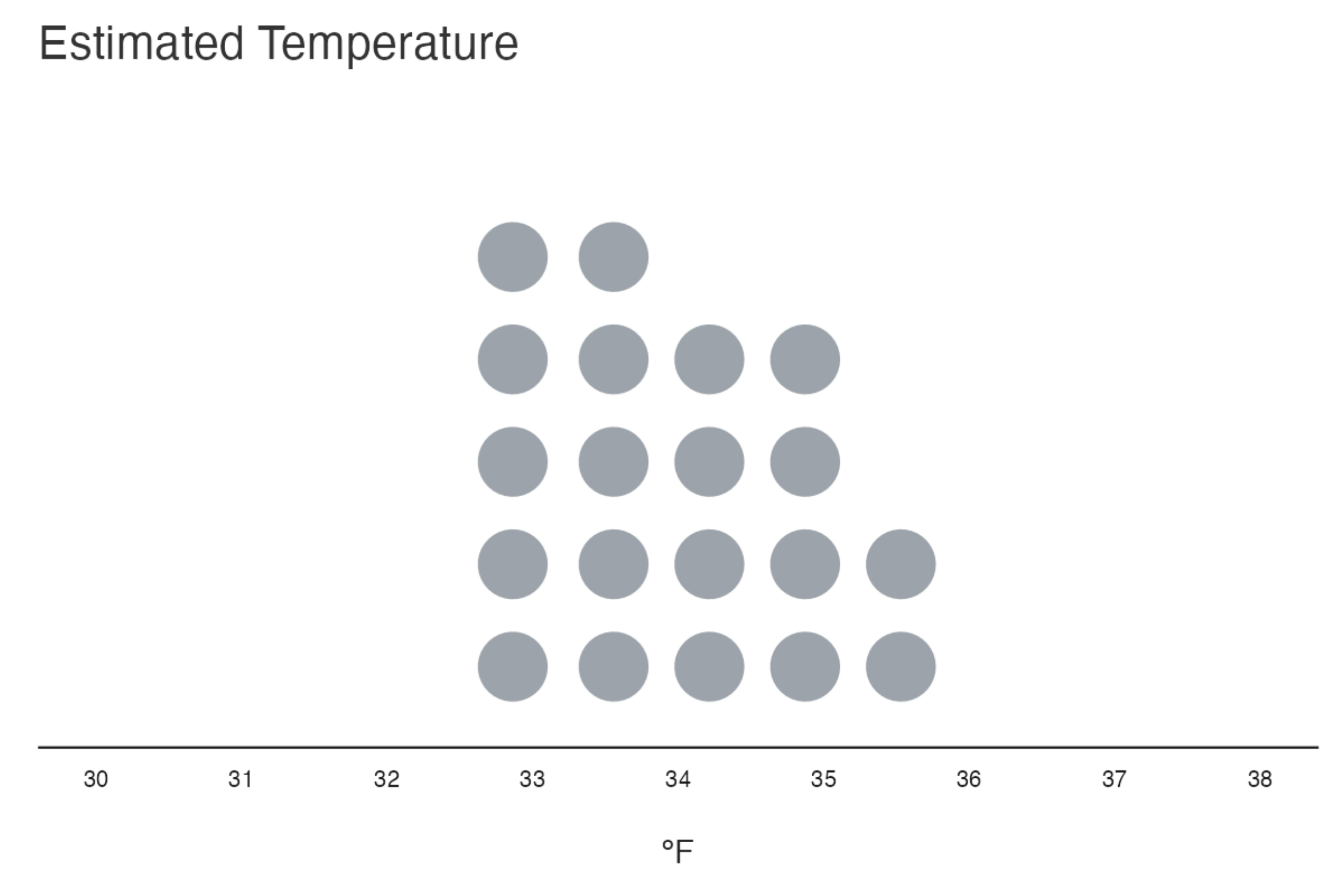}}
&  \cellcolor[HTML]{f2f2f2}\raisebox{-.85\totalheight}{\includegraphics[width=0.13\textwidth]{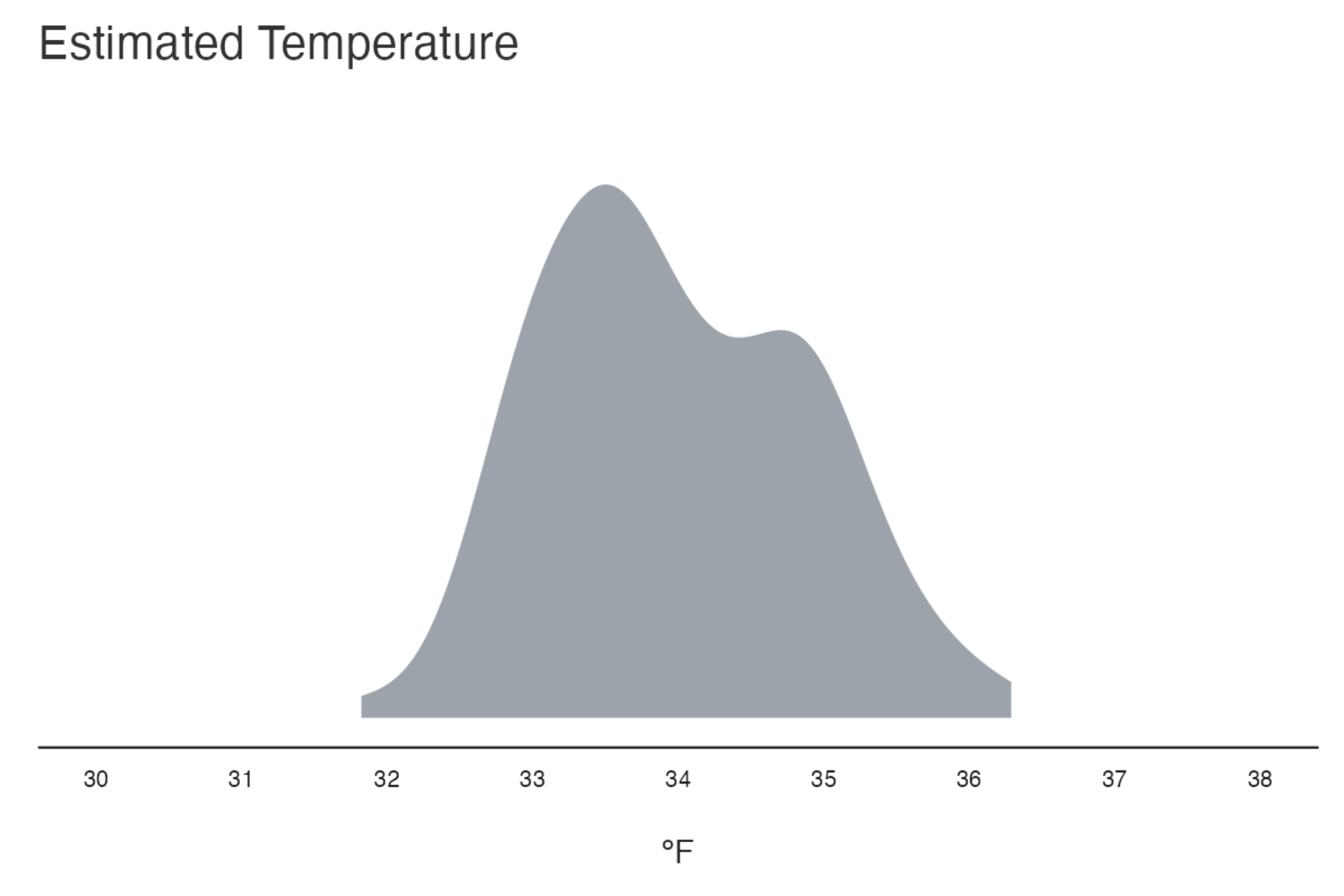}}
&  \raisebox{-.85\totalheight}{\includegraphics[width=0.13\textwidth]{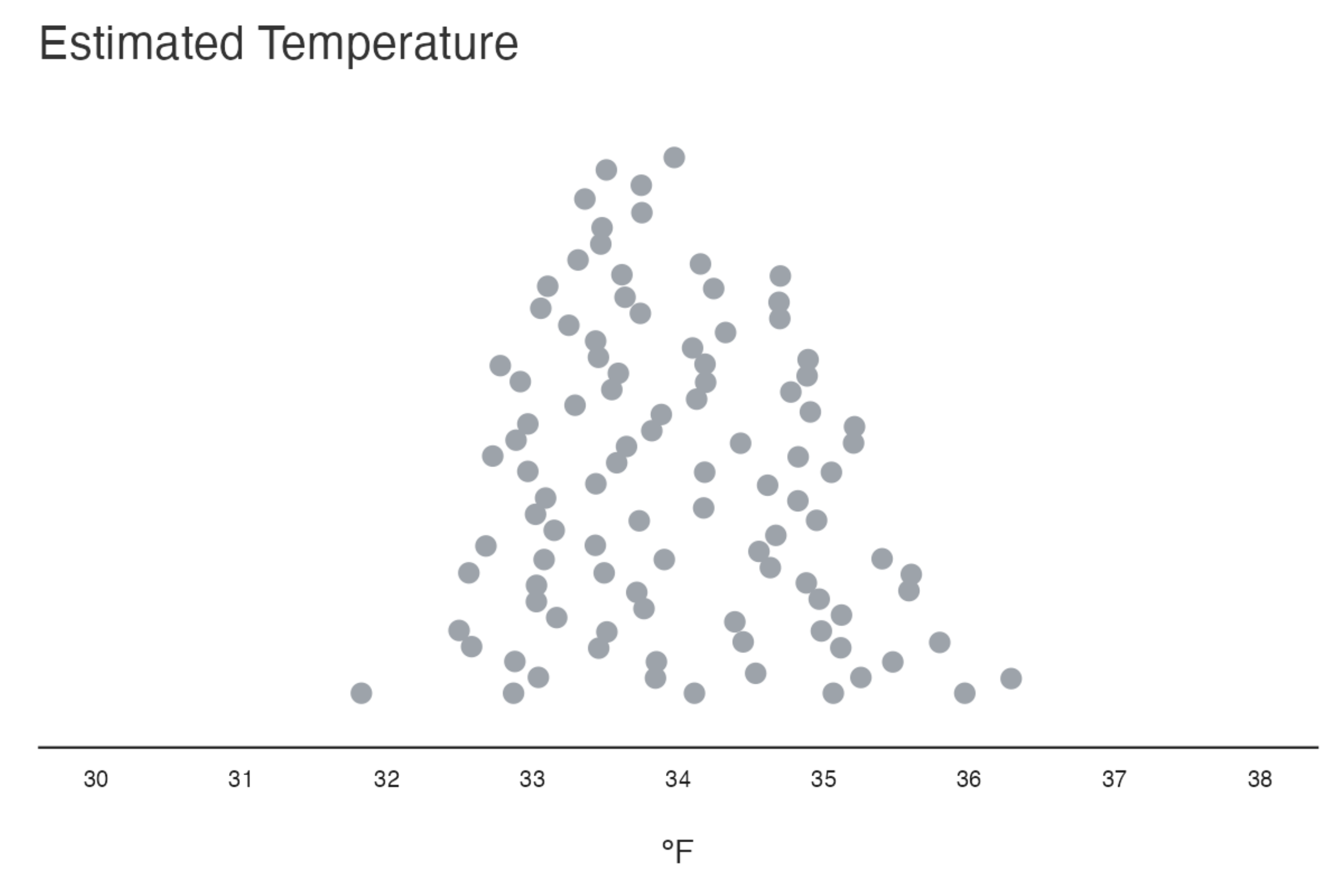}}
&  \raisebox{-.85\totalheight}{\includegraphics[width=0.13\textwidth]{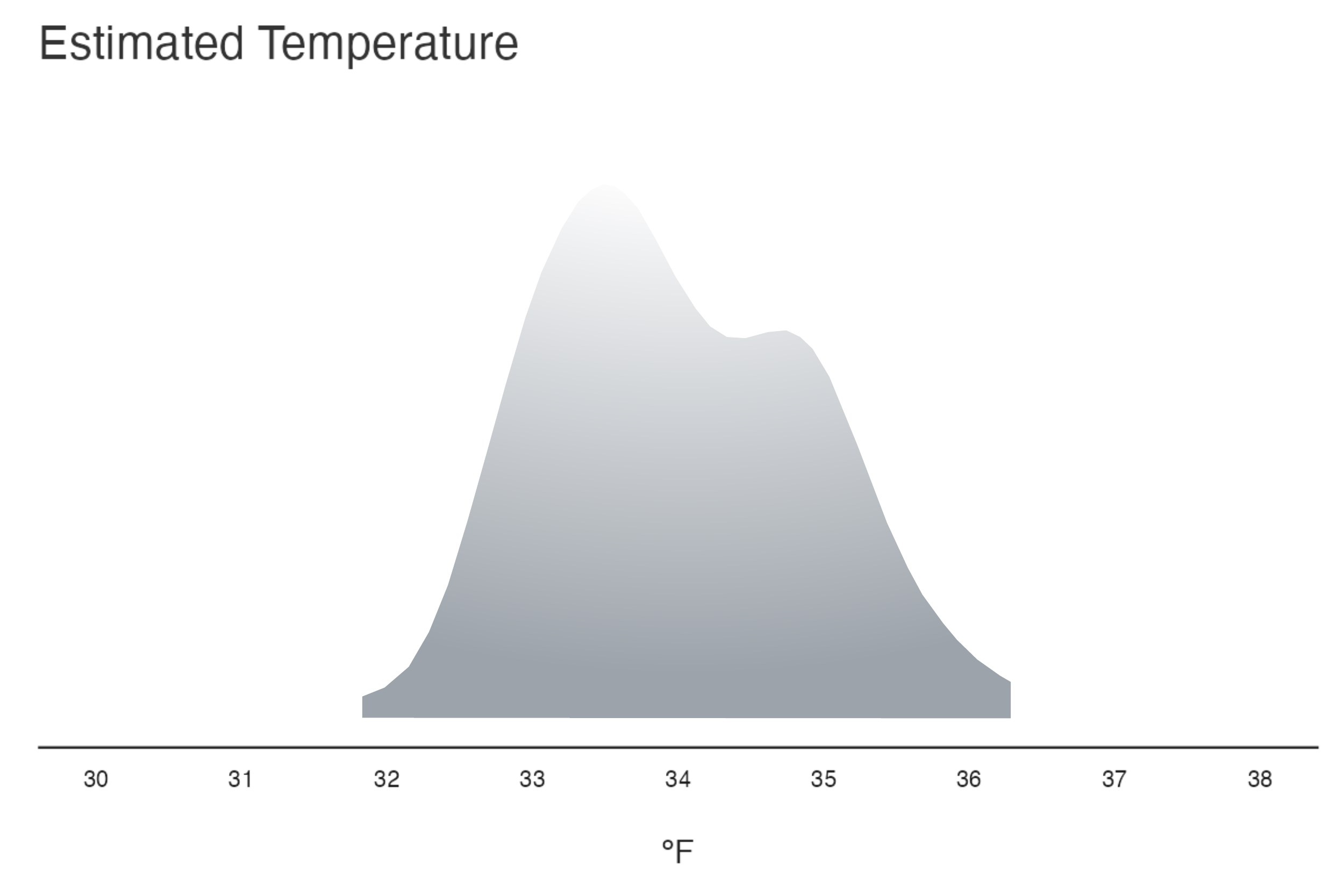}}
&  \raisebox{-.85\totalheight}{\includegraphics[width=0.13\textwidth]{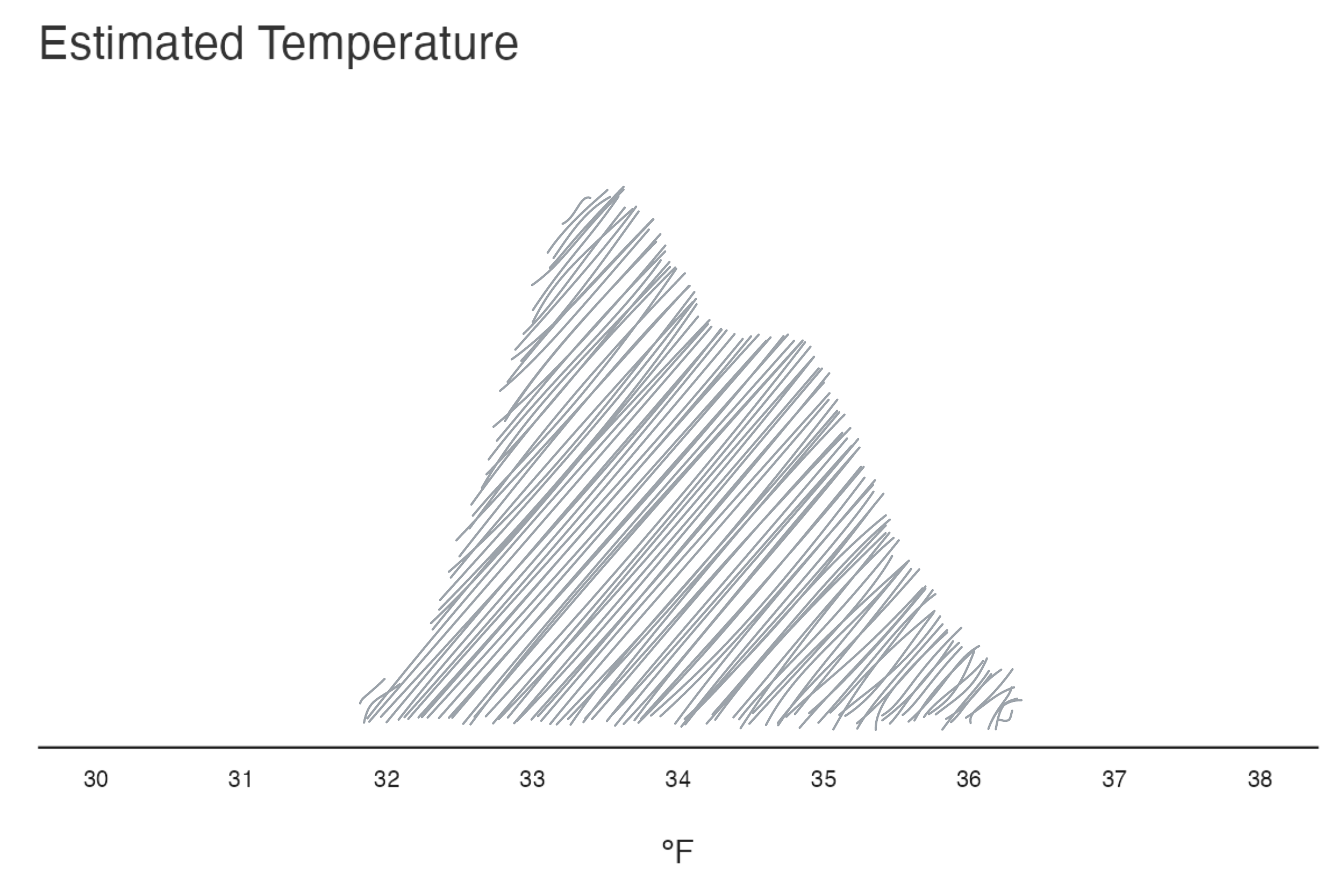}}
 \\ 
\hline
\textbf{Text (Sample)} & 
\cellcolor[HTML]{f2f2f2}The most likely temperature low tonight is [mean]ºF.  & 
The most likely temperature low tonight might be [mean]ºF. & 
\cellcolor[HTML]{f2f2f2}The most likely temperature low tonight might be around [mean]ºF. & 
The most likely temperature low tonight could potentially hover somewhere around [mean]ºF. & 
The most likely temperature low tonight \textit{could potentially} hover \textit{somewhere around} [mean]ºF.
 \\ 
\hline
\textbf{Speech} & 
\cellcolor[HTML]{f2f2f2}Concrete Text, No Modifications \newline (\textcolor{blue}{\href{https://osf.io/dnpa8?view\_only=78ba00b533b0485f9ba385e8f719693a}{MP3 Link}}) &   
Concrete Text, E1 Modifications \newline (\textcolor{blue}{\href{https://osf.io/tvpur?view\_only=78ba00b533b0485f9ba385e8f719693a}{MP3 Link}}) & 
Fuzzy Text, No Modifications \newline (\textcolor{blue}{\href{https://osf.io/rnf4u?view\_only=78ba00b533b0485f9ba385e8f719693a}{MP3 Link}}) &
\cellcolor[HTML]{f2f2f2}Fuzzy Text, E1 Modifications \newline (\textcolor{blue}{\href{https://osf.io/rwga8?view\_only=78ba00b533b0485f9ba385e8f719693a}{MP3 Link}}) &
Fuzzy Text, E1 Modifications and Question Contour \newline (\textcolor{blue}{\href{https://osf.io/nhv5y?view\_only=78ba00b533b0485f9ba385e8f719693a}{MP3 Link}}) \\ 
\hline
\end{tabular}
\end{table*}

While there has been work on basic representations of uncertainty for each mode, as detailed in Section \ref{section:related_work}, we expand that design space to include parameters of vagueness and uncertainty to better understand their influence on uncertainty communication and decision-making. We considered three main visualization types: beeswarm plots, quantile dot plots, and density plots, based on an informal survey of commonly used uncertainty visualizations and insights from previous research \cite{kay2016ish, fernandes2018uncertainty, padilla2021uncertain}. We varied components of these visualizations, including fuzziness, opacity, and arrangement~\cite{padillabook:2021}. We also included a variation in color, rendering the visualizations in orange, rather than gray (e.g., a density pot with an orange fill) \cite{smith2000users, tak2014color}. We analyzed a total of 21 visualizations, including two beeswarm plots, 10 density plots, five quantile dot plots, and four combinations of density and dot plots. Density plots are a common depiction of uncertainty, but we also investigated variations in frequency-based visualizations by adjusting the number of quantiles, color, and the overlap of other plot types.
This resulted in a set of 21 different visualizations illustrating the design space we were interested in exploring. 

For text variants, we considered the frequency and visual treatment of hedge words as well as the overall tone. Hedge words are key indicators of uncertainty in language, and this uncertainty may be emphasized by using uncertain visual appearances. We also included a ``colloquial'' condition, which used more informal language and everyday conversational expressions. We considered two visual treatments: italics and gray text. This resulted in a set of 15 text variants.
The speech design space was defined by variations in the duration of words and pauses, absolute pitch, and pitch contour. We selected these features as acoustic correlates of uncertainty found in earlier work (see Section \ref{rw_uncertainspeech}). We included variations in the content of the speech, resulting in 13 total variants.

To investigate these variants and identify a manageable design space of parameters for our experiments, the two pilot studies were conducted on Prolific \cite{palan2018prolific}, the first with 20 participants and the second with 40. Participants were fluent in English, had an approval rate above 95\%, and had normal color vision. They were introduced to the concepts of ``concrete'' and ``fuzzy'' information. 

Participants ranked sets of stimuli on a scale from concrete to fuzzy. The first pilot tested nine sets (4 visualization, 2 text, 3 speech), and the second pilot tested seven (4 visualization, 1 text, 2 speech). Since we could only make direct comparisons between stimuli in the same ranking set, we used the second pilot to compare new stimuli combinations, primarily narrowing the visualization sets. Participants initially completed a practice ranking where the items guided their rankings (e.g., ``This item should be ranked first'').
These rankings also served as a response quality check.

These pilot studies provided information to operationalize precision within individual communication modes. This method is not sufficient to standard uncertainty \textit{between} the different modes but instead assesses the impact of varying signals of uncertainty within each mode affect perceptions of fuzziness. These pilot studies ensured that we selected experimental stimuli with perceptible variations, which effectively varied the uncertainty signals in each mode. 

In selecting stimuli for the experiments, we carefully evaluated stimuli based on their average fuzziness rankings from the pilot studies. To do this, we placed the variants for each mode in their relative position on a spectrum and discussed the set of possible stimuli as a team. 
We chose stimuli that represented a range of most to least fuzzy, used diverse techniques in depicting uncertainty, were still within the realm of practical representations, and had basis in prior literature. 
This variety ensured a resulting set of stimuli with different precision levels and methods of representing fuzziness.

\subsection{Visualization Stimuli}

In the first pilot, participants perceived the quantile dot plot as fuzzier than the density plot. Their qualitative responses indicated they were unfamiliar with quantile dot plots. In the second pilot, after providing an explanation for quantile dot plots, participants ranked them as more concrete than density plots. The use of orange did not alter fuzziness rankings, but a lighter gray was ranked fuzzier than a darker gray. Gradient fills and sketchy styles heightened fuzziness as well, consistent with known visualization uncertainty encoding methods \cite{maceachren2012visual, franconeri2021science, wood2012sketchy}. 

For Experiment 1 (E1), we selected stimuli representative of the kinds of information displays often used in practice. We informally surveyed uncertainty visualizations in media and news reporting and incorporated feedback from a co-author with 15+ years of experience as a visualization designer. After this discussion, we selected the dark gray density plot, a common uncertainty representation, and the 20-quantile dot plot. While quantile dot plots required explanation to understand, prior work has indicated that they are an effective way to communicate uncertainty in a direct or frequency-based way \cite{fernandes2018uncertainty, kay2016ish, padilla2021uncertain}. These were representative of visualizations studied in prior work and displayed a difference in fuzziness ratings. For Experiment 2, we expand the range of fuzziness examined to include the beeswarm chart, a gradient density plot, and a density plot with a sketched style.

When creating these stimuli, we selected data from a normal distribution, using the \texttt{rnorm} function in R (v4.3.1) \cite{RComputing}. We selected 100 data points from a normal distribution with a standard deviation of 1 \cite{padilla2021uncertain}. True standard deviations ranged from 0.89 to 1.13. Averages ranged from 29.95 to 34.18ºF. This variation allowed a more ecologically valid set of distributions while still maintaining comparability; natural data will often not be perfectly normally distributed. Unless otherwise noted, all visualization stimuli were created using the \texttt{ggdist} package in R \cite{ggdist} with additional Figma modifications for the gradient and sketched designs. Stimuli code is included in supplementary materials. 

\subsection{Text Stimuli}
In text stimuli, more hedge words and visual emphases, like italics or gray color, increased perceived fuzziness. Stimuli with a colloquial tone received similar rankings as variants which used a more formal tone with a medium amount of hedge words. Interestingly, a passage with fewer but emphasized hedge words was ranked on average as fuzzy as one with more hedge words. 

In Experiment 1, we selected the variants without hedges and with medium hedging. In Experiment 2, we expanded this range to include minor, substantial, and italicized substantial hedging.
Text stimuli contained the most likely temperature (the rounded mean value), the middle 50\% range, the rounded minimum and maximum value, and a description of the distribution skew, calculated using \texttt{skewness} in the \texttt{moments} package in R \cite{moments}. 

\subsection{Speech Stimuli}

When ranking speech stimuli, increased and decreased pitches were viewed similarly, but a 10\% pitch change was slightly fuzzier than a 5\% change. However, 55\% of participants in the first pilot did not rank 0\% and 10\% pitch changes differently. Pitch contour was the most evident fuzzy indicator, followed by delay, then average pitch. Minor delays were perceived similarly to no delays, and significant delays matched E1 delays. Layering speech modifications, such as combining pitch with delay, consistently amplified fuzziness. 

For Experiment 1, we selected the default output with concrete text and a variant with delays, reduced rates, and minor pitch decreases with fuzzy text. The speech alterations applied only to number and/or hedge words. In Experiment 2, we added variants to further capture the impact of content and speech attributes, including concrete text with E1 modifications, fuzzy text with no modifications, and fuzzy text with E1 modifications and question contour. As some acoustic correlates of uncertainty are also correlates of emphasis, we used a lower pitch modification to use the acoustic cues as signaling uncertainty rather than emphasis; a higher pitch is a main correlate of emphasis \cite{LibermanPierrehumbert1984, KatzSelkirk2011}, but results for its relationship to confidence are varied \cite{JiangPell2015, SchererEtAl1973}. Speech stimuli were created using Google Speech Synthesis Markup Language (SSML)~\cite{googlessml}, a standardized markup language that allows adjustments in synthesized speech. Experiment 1 stimuli are shown in Table \ref{tab:e1_stimuli}, and Experiment 2 stimuli are shown in Table \ref{tab:e2_stimuli}.

\subsection{Discussion}

In this work, we mapped out a design landscape focused on the transmission of data uncertainty through various modes, examining how various techniques and visualization types influence the user's perception of uncertainty. This exploration of the design space provides two tangible contributions to the study of data uncertainty. 

\pheading{Signals of uncertainty.} Through these pilot studies, we identified and analyzed signals of uncertainty not evaluated in prior uncertainty research, such as text color and formatting. While prior work has outlined the mechanisms for emphasizing uncertainty in visual representations \cite{padillabook:2021}, it has not assessed how uncertain those representations \textit{appear} to readers. These pilot studies demonstrate one approach to comparing data representations and mechanisms of uncertainty. The findings from this design space exploration offer valuable guidance for professionals and further insight into the use of various techniques to emphasize uncertainty.

While the evaluation completed here was sufficient to select diverse, meaningful, and practical representations of uncertainty for the following experiments, future research should explore a more explicit and validated measure of how people interpret and react to uncertainty cues, including a clear mapping of perceptual thresholds across modes. Standardizing levels of fuzziness could provide more precise comparisons and a clearer understanding of how scaling uncertainty impacts decision-making across modes. 

\pheading{Implications for education.} Our study highlights the crucial role of providing explanations for visual representations of uncertainty, underscoring the need for strategies that teach individuals how to interpret these visualizations effectively. This insight highlights the important use of clear and accessible explanations alongside visualizations and opens several avenues for future research and development in educational strategies and tools.
Future research should focus on developing educational materials and methods that enhance comprehension of visual uncertainty, exploring which explanations are most effective for different audiences, and how these teachings can be integrated into various learning environments.

\section{Experiment Overview}
\label{section:expt_overview}

In Experiment 1, we compared six combinations of information modes (speech, text, visualization) and precision (concrete, fuzzy) for a $3 \times 2$ mixed-design study with the decision-making task. In Experiment 2, we examined the same three modes but with additional stimuli varying in precision. Understanding the unique capabilities of each mode is vital before we can assess how to merge them effectively in practical settings. This work provides a better understanding of how to approach situations where visualizations may not be effective and what alternatives could be employed in those instances. This dual focus not only maps out part of the space of unimodal representations but also guides the selection of the most appropriate communication mode based on the context.

The two experiments used a consistent decision structure from prior work \cite{joslyn2012uncertainty, padilla2021uncertain, nadav2009uncertainty, savelli2013advantages}. Participants were provided a data-driven task to decide whether to salt the road based on the evening's low-temperature forecast. This task was recognizable and relevant to participants, yet not so personal as to introduce subjective bias in their decisions. An objective, rational choice existed, determined by the cost and penalty values in the experiment.

Participants were presented with the following scenario: 
\noindent ``In cold weather, roads may need to be treated with salt to prevent icing. This salt treatment is costly but not as costly as damages caused by ice forming on roads. You are in charge of a road maintenance company contracted to treat the roads in a U.S. town with salt to prevent icing. It is your job to apply salt to the roads when the temperature is at or below 32ºF to prevent ice from forming.''

Participants were provided a fictional budget of \$12,000 to complete $12$ trials. Applying salt cost \$1,000, and the penalty for road freezing was \$3,000. Based on these values, the objectively rational choice was to salt when the chance of the temperature falling below 32ºF was at or above 33\%. Participants received a \$0.05 bonus for every \$1,000 left in their budget at the end of twelve trials. 

These experiments evaluated decision-making within this context, including confidence in the decisions and trust in the forecast. Participants' confidence levels shed light on their personal feelings during the decision-making process, as well as some insight into the perceived quality of their decision. As trust may alleviate the hesitations and doubts inherent in uncertain scenarios, collecting this information is key to a more comprehensive understanding of the trade-offs between different information modes. 

Confidence was assessed by asking participants to indicate the chance that their choice was correct, ranging from 50\% to 100\% \cite{tsai2008effects} (``How confident are you that you made the correct choice?''). We used a multi-item measure for ``trust,'' consisting of usefulness, clarity, and accuracy \cite{elhamdadi2022we, xiong2019examining,pandey2023you}. 

\subsection{Participants}
\label{section:e1e2_participants}

For both Experiments 1 and 2, we utilized the G*Power software \cite{faul2007g, faul2009statistical} for power analysis, aiming to achieve a power of $0.95$ with an alpha threshold of $0.05$. For Experiment 1, we estimated an effect size of $0.2$ based on pilot studies and previous research \cite{joslyn2012uncertainty, padilla2021uncertain}. For Experiment 2, we estimated an effect size of $0.15$ based on the results from Experiment 1.  Based on the power analysis, the necessary sample size was 105 participants for Experiment 1 and 129 participants per mode for Experiment 2.

Participants were recruited via the Prolific platform \cite{palan2018prolific}. Participants were omitted for irrelevant free-response answers. We also used an attention check for every trial, which queried the likelihood of a value falling outside the forecast's distribution range and thus with an expected likelihood of 0\%. If participants estimated a likelihood above 50\%, we discarded all their data. We set this threshold at 50\% rather than lower in order to accommodate potential challenges in probabilistic reasoning in text and speech scenarios.

To account for potential unusable responses, we recruited 20 more participants for each study than the power analysis indicated was necessary. 130 participants were recruited for Experiment 1, and 450 participants (150 per mode) were recruited for Experiment 2. The task took about 18 minutes, and participants received a payment of \$3.60. After exclusions, this resulted in 109 participants in Experiment 1. In Experiment 2, there were 128 participants in the speech condition, 132 in the text condition, and 136 in the visualization condition. Participants shared their age range and education; the demographic patterns are provided in supplementary materials. 

\subsection{Task Design}
\label{section:e1e2_design}

Experiments 1 and 2 used the same set of tasks. Participants were randomly assigned to a condition and then introduced to both the task and their assigned forecast type. This introduction explained potential decision outcomes and specifics of the assigned forecast. For those in the visualization group, there was a brief guide on interpreting the visual. Those in the speech group were informed they could replay the forecast multiple times if necessary. After this introduction, participants went through $12$ randomized trials. In each trial, participants viewed a forecast, made a decision, rated their confidence, estimated the evening's temperature and the likelihood of freezing, and answered the attention check.

Following the 12 trials, participants evaluated the forecast type they had been exposed to, assessing its clarity, accuracy, and usefulness. They also answered open-ended questions on their decision-making strategy and what they liked/disliked about the forecast. Participants also completed a four-question measure of graphical literacy \cite{okan2019using, burns2020evaluate, castro2021examining}. The survey concluded with demographic questions collecting age range and education level. Additionally, participants ranked their preferred way to receive information, choosing between visual (visualizations, images), written (books, articles), or audio (podcasts, radio broadcasts).

\subsection{Analysis}
\label{section:e1e2_analysis}

Methods and analyses were preregistered on OSF for both \href{https://osf.io/4ykep/?view_only=215f8f06f404404a9d187a9d76296d83}{Experiment 1} and \href{https://osf.io/8pz6n/?view_only=ed7df0a5f85c4e0f90515250b3c974bb}{Experiment 2}.
The analyses presented here deviate from preregistration; the initial plan included models with demographic variables that, upon further reflection and feedback, were not tied to specific hypotheses. As such, we adjusted our approach to evaluate only those models that incorporated directly relevant variables.

Both experiments investigated four dimensions of decision-making: the crossover temperature or turning point for the decision, the rationality of the decision, participants' confidence in their decision, and their overall trust in the information provided. 
Most of the results are based on mixed effects regression models using the \texttt{lmer} package in R, with model output generated by the \texttt{stargazer} package \cite{stargazer2022package}. Results for decision rationality are based on $\chi^2$ tests. 
All analysis materials, including full model results and statistical tests, are reported in supplementary materials.

\section{Experiment 1: Evaluation of Modes for Uncertainty Decision-making}

The goal of Experiment 1 was to examine \textbf{RQ1}: \textit{How does the mode of information presentation (speech, text, and visualization) affect decision-making with uncertain data?}, and how signals of uncertainty (precision) within each mode affect participant decisions.

\subsection{Hypotheses}
\label{section:e1_hypotheses}
People process information provided in text differently than speech \cite{FurnhamGunterGreen1990, KintschEtAl1975}. Using text, probabilities expressed with hedge words produce more conservative decisions than numeric probabilities \cite{nadav2009uncertainty}. For predictions using likelihood \textit{intervals}, text is more effective than visualizations \cite{savelli2013advantages}. However, visualizations typically outperform text in both challenging \cite{cheong2016evaluating} and simpler tasks \cite{mulder2020designing}.   Prior research found that visualizations allowing frequency reasoning, like quantile dot plots, resulted in better decisions and higher confidence than those that did not, such as density plots \cite{kay2016ish}.  For each of the decision dimensions listed in Section \ref{section:e1e2_analysis}, we assessed the effects of information mode (speech, text, or visualization) and precision (concrete vs. fuzzy).

\pheading{Determining crossover temperature.}
First, we evaluated the \textit{crossover temperature}. Crossover temperature is the turning point temperature at which participants were equally likely to salt or not salt. The \textit{optimal} crossover temperature is the point at which it becomes rational to salt the roads, determined by first identifying the freeze probability at which participants would ideally start salting, which is equivalent to the Cost:Penalty ratio (1:3 or 33\%.) If the provided data suggested a freezing likelihood above 33\%, salting the roads becomes the rational choice. We calculated the crossover temperature for each distribution by determining the mean value at which the distribution had a 33\% chance of 32ºF or below. Optimal crossover temperatures ranged from 32.3 to 32.6ºF. 

\noindent \textbf{H1a}: Visualization forecasts have a crossover temperature closest to optimal, followed by text, then speech. \textbf{H1b}: Concrete forecasts have a crossover temperature closer to optimal than fuzzy forecasts.

\pheading{Decision rationality.}
\noindent \textbf{H2a}: Visualization forecasts have more rational decisions than text or speech. No difference between text and speech. \textbf{H2b}: Concrete forecasts have more rational decisions than fuzzy forecasts.

\pheading{Decision confidence.}
\noindent\textbf{H3a}: Visualization forecasts have the highest confidence ratings, followed by text, then speech. \textbf{H3b}: Concrete forecasts have higher confidence ratings than fuzzy forecasts.

\pheading{Trust in forecast.}
Finally, we examined trust ratings. As trust is a complex process, we examined three dimensions of trust: usefulness, clarity, and accuracy \cite{elhamdadi2022we, xiong2019examining}. Visual representations afford probabilistic reasoning more than their text and speech counterparts. This increased information may increase clarity and perceived reliability. A concrete representation may also increase clarity and reliability in comparison to a fuzzy representation.

\noindent \textbf{H4a}: Visualization forecasts have the highest trust ratings, followed by text, then speech. \textbf{H4b}: Concrete forecasts have higher trust ratings than fuzzy forecasts.

\subsection{Results}
\label{section:e1_results}

In general, participants performed well in the decision-making task, with  \$2,531 remaining on average. Most participants played the speech stimuli only once (72\% of trials). The results indicated that both visualizations and text supported rational decision-making, but text led to lower confidence ratings. Speech, while associated with riskier decisions, led to higher trust ratings than the other two modes.
Precision did not appear to have an impact on decisions.

\subsubsection{H1: Decision Crossover Temperature}
\textbf{H1a} and \textbf{H1b} were evaluated with logistic mixed effects models predicting the binary salting decision: ``Do not salt'' (0) or ``Salt'' (1). Consistent with prior work, we used these predictions to calculate the deviations from optimal crossover temperatures \cite{padilla2021uncertain}. These calculations accounted for variations between distributions, as they were based on the distance from the mean of the distribution to the optimal crossover temperature for each distribution. 

We compared models using an ANOVA test for model selection. The optimal model ($p = 0.014$) included a random effect of participant and fixed effects of the difference between the average temperature and the optimal crossover, and the mode of information. 

Table \ref{tab:e1_crossovertemps} illustrates the differences from the optimal crossover temperature. For speech and text stimuli, people were more likely to be risky than conservative, indicated by the negative values in each cell, and vice versa for visualization stimuli. Perfectly optimal decision-making would result in a value of 0.

We found partial support for \textbf{H1a}. Visualization forecasts had a crossover temperature closer to optimal than speech, resulting from an increased likelihood of salting the roads (2.7 - 7.6x more likely, $p = 0.004$). There was a minor difference between text and speech, with text decisions being more likely to salt (1.6 - 4.6x more likely, $p = 0.055$). There was no difference between text and visualization ($p = 0.324$). Including precision did not improve model performance, so we do not find support for \textbf{H1b}. The crossover temperature pattern was similar to the binary salting decisions. 

\begin{table}[ht]
    \centering
\renewcommand{\arraystretch}{1.25}
\arrayrulecolor{lightgray}
\caption{Difference from optimal crossover temperature. Darker gray indicates a more optimal crossover temperature.}
\label{tab:e1_crossovertemps}
\begin{tabular}
{|l|r|r|}
\hline
&  \textbf{Concrete} & \textbf{Fuzzy} \\ 
\hline
\textbf{Speech} & 
\cellcolor[HTML]{FFFFFF}$-$0.517 & 
\cellcolor[HTML]{E9E9E9}$-$0.404 \\ 
\hline
\textbf{Text} & 
\cellcolor[HTML]{B7B7B7}$-$0.151 &
\cellcolor[HTML]{A1A1A1}$-$0.038
\\
\hline
\textbf{Visualization} & 
\cellcolor[HTML]{9F9F9F} 0.026 &
\cellcolor[HTML]{B5B5B5} 0.139
\\
\hline
\end{tabular}

\end{table}

\subsubsection{H2: Decision Rationality}

The relatively close match of crossover temperatures to optimal values suggests a strong pattern of rational decision-making. Participants' decisions were grouped into three categories: rational, conservative, or risky. Conservative decisions salted the roads despite the probability of freezing less than 33\%. Risky decisions did not salt despite the probability of freezing greater than 33\%. We examined \textbf{H2a} and \textbf{H2b} about decision rationality using $\chi^2$ tests.

\begin{figure}[ht]
    \centering
    \includegraphics[width = 0.99\linewidth]{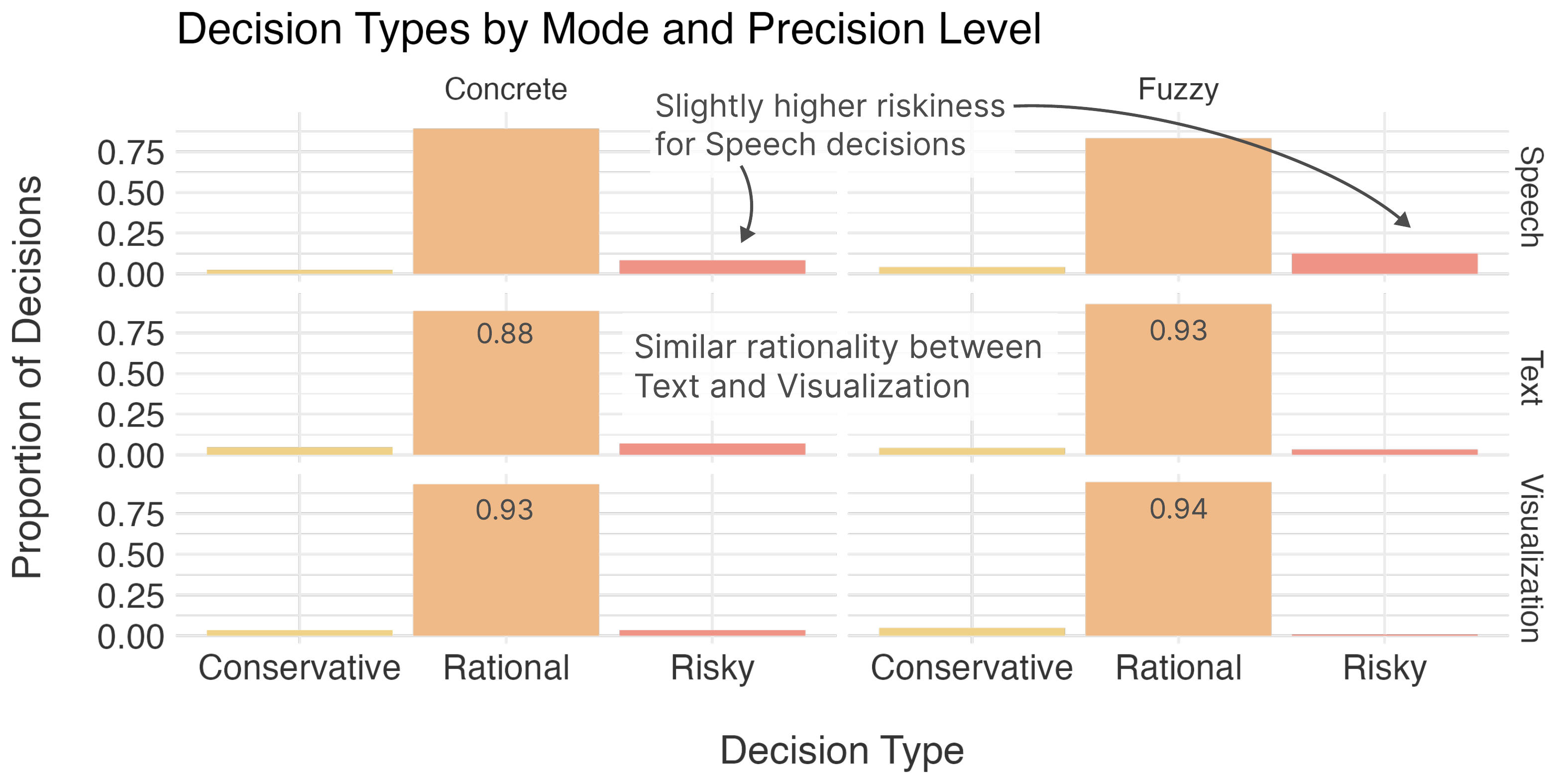}
    \caption{Proportion of decision types for each condition in Experiment 1. Overall, decisions were mostly rational. Speech was the least rational mode, with a greater proportion of risky decisions.}  
    \label{fig:e1_rationality}
\end{figure}

Decisions based on visualization forecasts were more often rational compared to the more risky decisions based on speech forecasts, providing partial support for \textbf{H2a} ($\chi^2 = 30.1$, $p < 0.01$), as decisions with text forecasts exhibited intermediate rationality. \textbf{H2b} was not supported ($\chi^2 = 1.16 $, $p = 0.559$), similar to with results from \textbf{H1}. Figure \ref{fig:e1_rationality} shows the decision types by condition.

\subsubsection{H3: Decision Confidence}

\begin{figure}[ht]
    \centering
    \includegraphics[width = 0.99\linewidth]{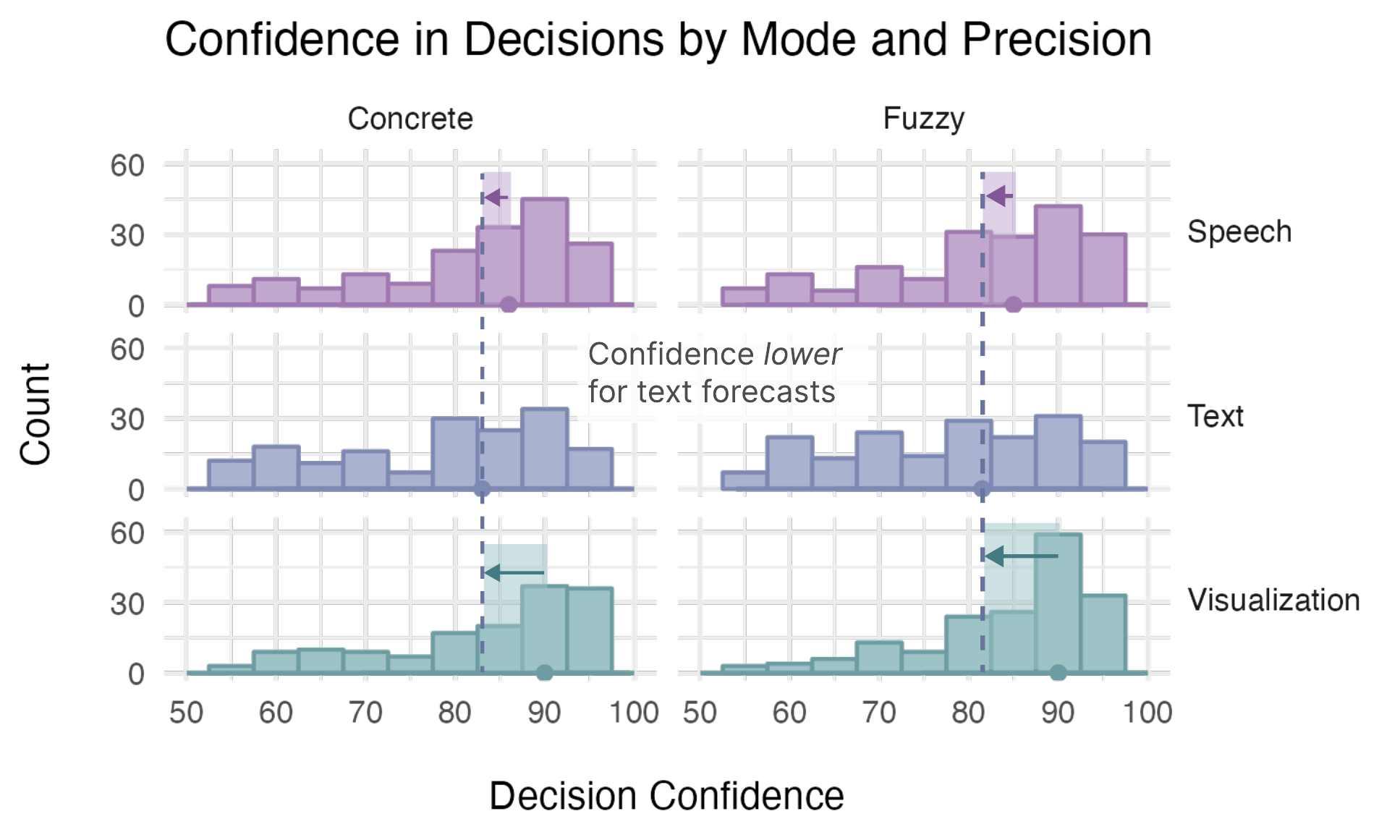}
    \caption{Experiment 1 confidence ratings ranged from 50 to 100. Confidence was lower for text than for visualization forecasts.
    }
    \label{fig:e1_confidence}
\end{figure}

Participant confidence ratings are shown in Figure \ref{fig:e1_confidence}.
We compared relevant models using an ANOVA. The optimal model ($p = 0.014$) included a random effect for participant and fixed effects of the difference between mean temperature and the optimal crossover, decision rationality, and mode of information. Including precision did not improve model performance; we did not find support for \textbf{H3b}. We found partial support for \textbf{H3a} - visualization forecasts produced higher confidence ratings compared to text ($p = 0.005$) by 7-15\%. There was no notable difference between speech and text ($p = 0.131$) nor speech and visualization ($p = 0.188$).

\subsubsection{H4: Trust in Forecast}

Finally, we conducted an analysis of trust ratings.  We used the same ANOVA process to select the optimal model.  The final model ($p = 0.003$) included the rationality of the decision and the mode of the forecast.  Precision (concrete vs. fuzzy) did not significantly improve the model ($p = 0.984$), so it was not included as a factor. Thus, we did not find support for \textbf{H4b}. The results of this model \textit{contradicted} \textbf{H4a}: speech forecasts were the most trusted, with 9-17\% higher ratings than text forecasts ($p < 0.01$) and 6-14\% higher ratings than visualization forecasts ($p = 0.013$).

\begin{figure}[ht]
    \centering
    \includegraphics[width = 0.99\linewidth]{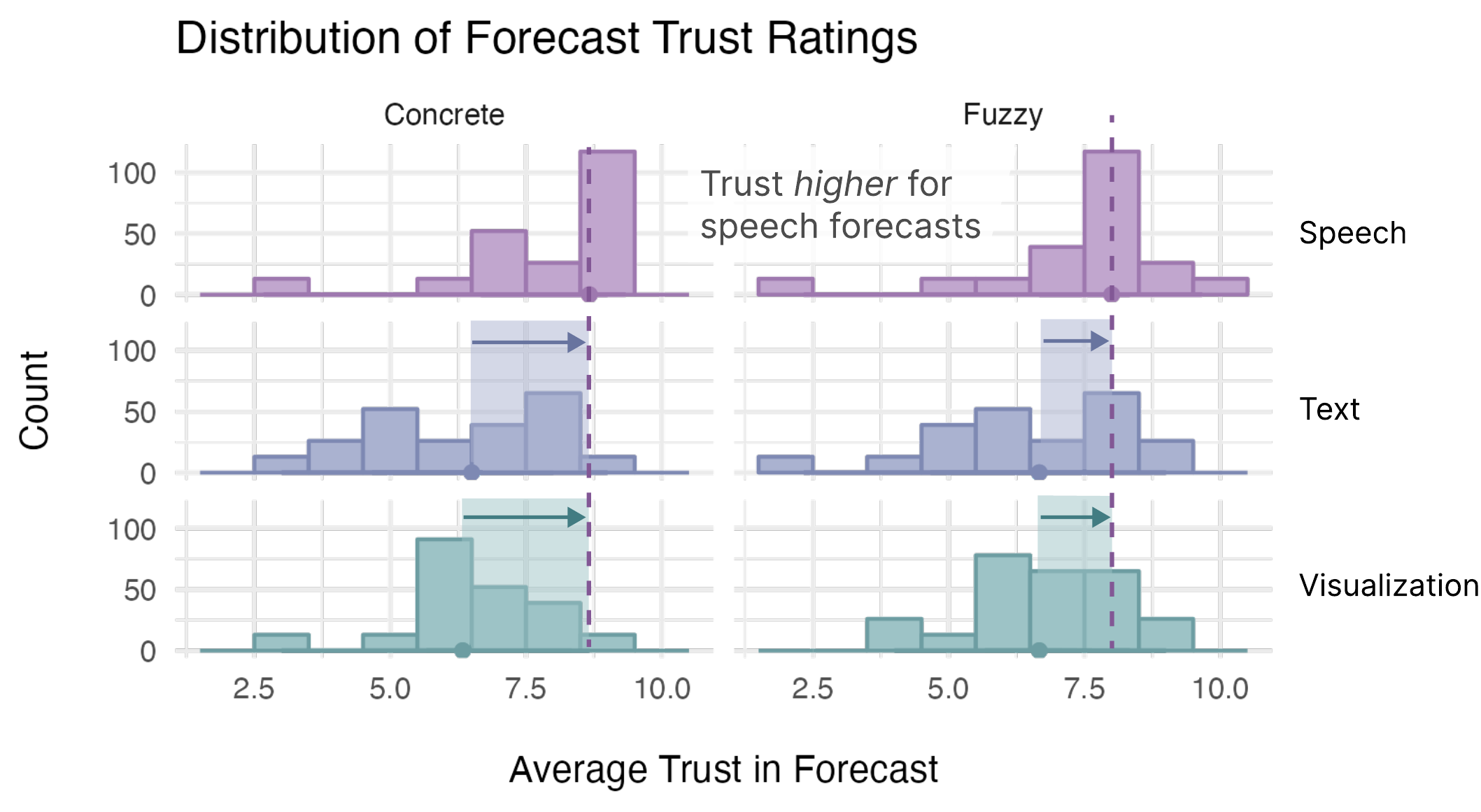}
    \caption{Experiment 1 trust ratings. Participants trusted speech the most in comparison to both text and visualization.}
    \label{fig:e1_trust}
\end{figure}

\subsection{Discussion}

The findings of this study offer insights into how participants interpret uncertain information presented in different modes.

\pheading{High trust in speech.} Speech forecasts received the highest trust ratings (Figure \ref{fig:e1_trust}); speech also led to a decreased likelihood to salt the roads and thus riskier decisions than visualizations. This combination of results presents a complex trade-off: while speech was perceived as more trustworthy, it did not guide participants as accurately as other modes. When supported by other modes, speech may improve trust without sacrificing decision-making.

\pheading{Lower confidence with text.} Text forecasts were a strong mode for conveying outcomes and uncertainty. Decisions made using text were on par with visualizations in terms of crossover temperature (Table \ref{tab:e1_crossovertemps}) and rationality (Figure \ref{fig:e1_rationality}). However, participants were less confident with text-based forecasts than with visualizations. Given that text can act as a strong support to decision-making, it may be best used when integrated with other communicative mediums.

\section{Experiment 2: Effect of Mode Design Factors on Uncertainty Decision-making}

We aimed to further explore variations within each mode to better encompass the design space outlined in Section \ref{section:design_space}. In Experiment 2, we asked - \textbf{RQ2}: \textit{How does decision-making change when signs of uncertainty are intensified within each mode of communication?}

\subsection{Hypotheses}
Experiment 2 investigated the same decision dimensions described in Section \ref{section:e1e2_analysis}. We compared precision levels of stimuli within each mode. Hypotheses are based on trends observed in Experiment 1. 

\pheading{Determining crossover temperature.}
\textbf{H5a-b}: Crossover temperature is closer to optimal as \textbf{(a) speech} and  \textbf{(b) text} forecasts increase in fuzziness. \textbf{H5c}: Crossover temperature is further from optimal as \textbf{visualization} forecasts increase in fuzziness.

\pheading{Decision rationality.} 
\textbf{H6a-b}: Decisions are \textit{more} frequently rational as \textbf{(a) speech} and  \textbf{(b) text} forecasts increase in fuzziness. \textbf{H6c}: Decisions are \textit{less} frequently rational as  \textbf{visualization} forecasts increase in fuzziness.

\pheading{Decision confidence.}
\textbf{H7a-c}: Confidence in the decision increases as \textbf{all three modes} increase in fuzziness.

\pheading{Trust in forecast.}
\textbf{H8a-c}: Trust in the forecast decreases \textbf{all three modes} increase in fuzziness.

\subsection{Results}
\label{section:e2_results}

As in Experiment 1, participants generally performed well at the task, with \$1,737 remaining on average. Most participants played the speech stimuli only once (79\% of trials). Figures for rationality, confidence, and trust can be found in supplemental materials.

\subsubsection{H5: Determining Crossover Temperature}

As in Experiment 1, we used ANOVA for model selection. Including precision as a factor did not improve model performance for the speech ($p = 0.957$), text ($p = 0.884$), or visualization models ($p = 0.133$). We did not find support for an effect of precision on the likelihood of salting and crossover temperatures (\textbf{H5a-c}).

Participants make riskier decisions in this experiment than in Experiment 1, as almost all crossover temperatures were positive (Table \ref{tab:e2_crossovertemps}). This may be related to the differences in the participant pool or specific distributions, as they differed from Experiment 1. Additionally, decisions were closer to optimal with speech forecasts than in the prior experiment, especially in comparison to text forecasts.

\begin{table*}[ht]
    \centering
\renewcommand{\arraystretch}{1.25}
\caption{Difference from optimal crossover temperature. Darker gray indicates a more optimal crossover temperature.
}
\label{tab:e2_crossovertemps}
\begin{tabular}
{|l|r|r|r|r|r|}
\hline
& \textbf{Most concrete} & \textbf{Somewhat concrete} & \textbf{Mixed} & \textbf{Somewhat fuzzy} & \textbf{Most fuzzy} \\ 
\hline
\textbf{Speech} & \cellcolor[HTML]{A5A5A5}0.054 & \cellcolor[HTML]{9F9F9F}0.029 & \cellcolor[HTML]{9B9B9B}$-$0.011 & \cellcolor[HTML]{ADADAD}0.089 & \cellcolor[HTML]{B3B3B3}0.113 \\
\hline
\textbf{Text} & \cellcolor[HTML]{DADADA}0.281 & \cellcolor[HTML]{FFFFFF}0.439 & \cellcolor[HTML]{F1F1F1}0.380 & \cellcolor[HTML]{E7E7E7}0.338 & \cellcolor[HTML]{EEEEEE}0.367 \\
\hline
\textbf{Visualization} & \cellcolor[HTML]{9C9C9C}$-$0.015 & \cellcolor[HTML]{BBBBBB}0.148 & \cellcolor[HTML]{B2B2B2}0.110 & \cellcolor[HTML]{E1E1E1}0.312 & \cellcolor[HTML]{C9C9C9}0.209 \\
\hline
\end{tabular}

\end{table*}

\subsubsection{H6: Decision Rationality}

Decisions were again categorized as rational, conservative, or risky and evaluated using $\chi^2$ tests. There were no significant differences in the rationality of decisions based on precision for speech ($\chi^2 = 4.07$, $p = 0.850$) or text stimuli ($\chi^2 = 5.40$, $p = 0.714$); there was no support for \textbf{H6a-b}. However, there was a significant effect of precision on rationality with visualization stimuli ($\chi^2 = 15.9$, $p = 0.0443$).  Participants made more conservative decisions in the somewhat fuzzy condition (density plot with gradient) than in the most concrete condition (quantile dot plot). However, there was no consistent effect of precision, providing unclear support for \textbf{H6a}.

\subsubsection{H7: Decision Confidence}

As in Experiment 1, we examined participants' confidence in their decisions. We used ANOVA testing for model selection;  Including precision as a factor did not improve model performance for the speech ($p = 0.656$), text ($p = 0.244$), or visualization model ($p = 0.509$). As such, we did not find support for \textbf{H7a-c}. We did not find a connection between decision confidence and forecast fuzziness. 

\subsubsection{H8: Trust in Forecast}

Finally, we conducted an analysis of participants' trust ratings for the forecast. Precision did not significantly improve model performance for speech ($p = 0.133$) or visualization ($p = 0.080$); we did not find support for \textbf{H8a} or \textbf{H8c}. For text stimuli, the best model ($p = 0.012$) included decision rationality and stimulus precision. Most concrete was the most trusted, while most fuzzy was less trusted than other variants.  While the presence of hedge words influenced trust, there was no evidence for a continuous pattern based on the hedge frequency. However, emphasizing these hedge words by italicizing them \textit{decreased} trust. This finding is contrary to \textbf{H8b}.

\subsection{Effects of Mode}

The Experiment 1 conditions were also present in Experiment 2, so we ran the analyses from Experiment 1 on the Experiment 2 data to test the consistency of findings across the experiments. We consider concrete and fuzzy categories from Experiment 1; results largely support our initial findings, with variation in decision quality. 

We did \textit{not} find that participants were less likely to apply salt with speech forecasts than with text ($p = 0.209$) or visualizations ($p = 0.884$). Likewise, we did not find that speech led to riskier decision-making ($\chi^2 = 7.00$, $p = 0.136$). We did observe a difference by precision; fuzzy forecasts across modes led to more conservative decisions than concrete ($\chi^2 = 7.22$, $p = 0.027$).

We again found a lower confidence level for text, this time in comparison to speech ($p = 0.002$) rather than visualization. ($p = 0.186$). Despite relatively similar decision quality, text again received lower confidence ratings. We also found that speech forecasts had higher trust ratings than visualization ($p = 0.043$).

\subsection{Discussion}
When comparing the decisions made with these variants within each mode, we found little effect of fuzziness. There was some effect of hedging on trust and a connection between visualization fuzziness and conservative decision-making. However, these results were not consistent across variants and did not provide a clear effect of precision. These results may reflect a true lack of effect of precision on decision-making; they may also be due to the selected manipulations of fuzziness; more extreme variants (e.g., significant pitch changes) may result in different effects. We constrained our stimuli set to realistic variants for ecological validity. 

Between the two experiments, we found that, despite similar success at the task, participants were riskier in Experiment 1 and more conservative in Experiment 2. This may explain some of the differences between the results for each study; the impact of fuzziness may vary based on how participants were approaching the task. However, the findings in trust and confidence were consistent.

\section{Discussion and Future Work}

Our exploration of speech, text, and visualization for conveying data uncertainty revealed several variations in decision-making. 

\noindent\textbf{RQ1}: \textit{Comparison of Modes - How does the mode of information presentation affect decision-making with uncertain data?} Visualization and text forecasts supported rational decision-making, while speech introduced a slightly increased level of risk. Decisions using text received lower confidence ratings than visualizations. Speech received higher trust ratings than the other modes.

\noindent \textbf{RQ2}: \textit{Comparison of Precision - How does decision-making change when indicators of uncertainty are intensified within each mode of communication?} Increasing indicators of uncertainty did not have a measurable impact on decisions. The confidence and trust findings from Experiment 1 were supported in Experiment 2.

\subsection{Limitations and Future Directions}

\pheading{Explore multimodal representations.} 
In this work, we uncovered key trade-offs for individual modes. Speech representations can lead to higher levels of trust but may also increase irrational decision-making. Text representations provide support for rational decisions but lower confidence in these decisions. Visualizations are useful for decision-making overall but can be difficult to interpret and require additional explanation. While we did not investigate multimodal representations, our results suggest a need for further research on how people make decisions when presented with uncertainty data in multimodal contexts. 

\pheading{Expand the design space.}
The primary focus of our work was to understand the impact of different modes on the perception of data uncertainty. By keeping elements like visualization style, font, and speaker characteristics constant, the study aimed to isolate the effects of the mode of information itself. Exploring additional elements of each mode would substantially increase the complexity of the study design space and is outside the scope of our study. Future research could explore how additional design factors such as the speaker's (perceived) gender or accent~\cite{Giles1970, MulacRudd1977, StewartRyanGiles1985} might affect the perception of uncertainty. 

\pheading{Explore additional uncertainty tasks.} 
Our work considered one type of decision-making task as described in \cite{padilla2018decision}. While other work in uncertainty has explored other kinds of decisions (e.g., transportation \cite{kay2016ish}), these studies typically examine only one mode of information rather than comparing different representations. Future studies could diversify scenarios to include personal experiences with related activities, such as living in cold climates or financial decisions (e.g., funding a startup) to better understand how modes affect different categories of decision-making.

\pheading{Supporting accessibility.} Future research should address how to support accessibility when designing multimodal interfaces for data uncertainty. While the current study did not specifically delve into evaluation based on accessibility, designing equitable interfaces is an important research direction to consider and one that these findings support; accessible interfaces will provide flexibility for users to choose modes that best align with their abilities and preferences.

\section{Conclusion}

Effectively communicating and understanding data uncertainty is essential for making everyday decisions and in domains such as finance, healthcare, and policy-making. This work explores how different modes - speech, text, and visualization, can be used to convey uncertainty. From hedge words and prosody variants to quantile dot and density plots, we explore a design space within each mode. Through two crowdsourced studies, we identify how the mode of information influences decision-making and trust when reasoning with uncertain data. While visualization and text best support decision-making, users found speech more trustworthy and had lower confidence in decisions made with text. These findings suggest future research directions that explore a more nuanced approach to uncertainty communication.

\bibliographystyle{eg-alpha-doi} 
\bibliography{0_bib}       

\newcommand{\etalchar}[1]{$^{#1}$}
\begin{thebibliography}{\uppercase{KKGMH20}}

\bibitem[BH05]{BlankenshipHoltgraves2005}
\textsc{Blankenship K.~L., Holtgraves T.}:
\newblock The {R}ole of {D}ifferent {M}arkers of {L}inguistic {P}owerlessness in {P}ersuasion.
\newblock \emph{Journal of Language and Social Psychology 24}, 1 (2005), 3--24.

\bibitem[BHJ{\etalchar{*}}14]{Bonneau2014OverviewAS}
\textsc{Bonneau G.-P., Hege H.-C., Johnson C.~R., de~Oliveira~Neto M.~M., Potter K.~C., Rheingans P., Schultz T.}:
\newblock Overview and {S}tate-of-the-{A}rt of {U}ncertainty {V}isualization.
\newblock In \emph{Scientific Visualization: Uncertainty, Multifield, Biomedical, and Scalable Visualization} (London, 2014), Hansen C.~D., Chen M., Johnson C.~R., Kaufman A.~E., Hagen H., (Eds.), Springer, pp.~3--27.

\bibitem[BW95]{BrennanWilliams1995}
\textsc{Brennan S.~E., Williams M.}:
\newblock The {F}eeling of {A}nother's {K}nowing: {P}rosody and {F}illed {P}auses as {C}ues to {L}isteners {A}bout the {M}etacognitive {S}tates of {S}peakers.
\newblock \emph{Journal of Memory and Language 34} (1995), 383--398.

\bibitem[BXF{\etalchar{*}}20]{burns2020evaluate}
\textsc{Burns A., Xiong C., Franconeri S., Cairo A., Mahyar N.}:
\newblock How to {E}valuate {D}ata {V}isualizations across {D}ifferent {L}evels of {U}nderstanding.
\newblock In \emph{2020 IEEE Workshop on Evaluation and Beyond-Methodological Approaches to Visualization (BELIV)} (Piscataway, NJ, 2020), IEEE, pp.~19--28.

\bibitem[CBD09]{CramerEtAl2009}
\textsc{Cramer R.~J., Brodsky S.~L., DeCoster J.}:
\newblock Expert {W}itness {C}onfidence and {J}uror {P}ersonality: {T}heir {I}mpact on {C}redibility and {P}ersuasion in the {C}ourtroom.
\newblock \emph{Journal of the American Academy of Psychiatry and the Law Online 37}, 1 (2009), 63--74.

\bibitem[CBK{\etalchar{*}}16]{cheong2016evaluating}
\textsc{Cheong L., Bleisch S., Kealy A., Tolhurst K., Wilkening T., Duckham M.}:
\newblock Evaluating the {I}mpact of {V}isualization of {W}ildfire {H}azard upon {D}ecision-making {U}nder {U}ncertainty.
\newblock \emph{International Journal of Geographical Information Science 30}, 7 (2016), 1377--1404.

\bibitem[CDRS03]{ChattopadhyayEtAl2003}
\textsc{Chattopadhyay A., Dahl D.~W., Ritchie R.~J., Shahin K.~N.}:
\newblock Hearing {V}oices: The {I}mpact of {A}nnouncer {S}peech {C}haracteristics on {C}onsumer {R}esponse to {B}roadcast {A}dvertising.
\newblock \emph{Journal of Consumer Psychology 13}, 3 (2003), 198--204.

\bibitem[CJTD21]{CapurroEtAl2021}
\textsc{Capurro G., Jardine C.~G., Tustin J., Driedger M.}:
\newblock Communicating {S}cientific {U}ncertainty in a {R}apidly {E}volving {S}ituation: {A} {F}raming {A}nalysis of {C}anadian {C}overage in {E}arly {D}ays of {COVID-19}.
\newblock \emph{BMC Public Health 21}, 1 (2021), Article 2181.

\bibitem[CQHP21]{castro2021examining}
\textsc{Castro S.~C., Quinan P.~S., Hosseinpour H., Padilla L.}:
\newblock Examining {E}ffort in 1d {U}ncertainty {C}ommunication {U}sing {I}ndividual {D}ifferences in {W}orking {M}emory and {NASA-TLX}.
\newblock \emph{IEEE Transactions on Visualization and Computer Graphics 28}, 1 (2021), 411--421.

\bibitem[DBRS08]{DurikEtAl2008}
\textsc{Durik A.~M., Britt M.~A., Reynolds R., Storey J.}:
\newblock The {E}ffects of {H}edges in {P}ersuasive {A}rguments: {A} {N}uanced {A}nalysis of {L}anguage.
\newblock \emph{Journal of Language and Social Psychology 27}, 3 (2008), 217--234.

\bibitem[DBY18]{DesenderBoldtYeung2018}
\textsc{Desender K., Boldt A., Yeung N.}:
\newblock Subjective {C}onfidence {P}redicts {I}nformation {S}eeking in {D}ecision {M}aking.
\newblock \emph{Psychological Science 29}, 5 (2018), 761--778.

\bibitem[DH01]{deHaan2001}
\textsc{De~Haan F.}:
\newblock The {R}elation between {M}odality and {E}videntiality.
\newblock \emph{Linguistische Berichte 9} (2001), 201--16.

\bibitem[DM20]{ding:2020}
\textsc{Ding Q., Millet B.}:
\newblock Visualizing {U}ncertainty in {W}eather {F}orecasts.
\newblock \emph{Proceedings of the Human Factors and Ergonomics Society Annual Meeting 64} (12 2020), 1064--1068.
\newblock \href {https://doi.org/10.1177/1071181320641255} {\path{doi:10.1177/1071181320641255}}.

\bibitem[Dra95]{draper:1995}
\textsc{Draper D.}:
\newblock Assessment and {P}ropagation of {M}odel {U}ncertainty.
\newblock \emph{Journal of the Royal Statistical Society, Series B (Methodological) 57} (1995), 45--97.
\newblock \href {https://doi.org/10.2307/2346087} {\path{doi:10.2307/2346087}}.

\bibitem[EGKX22]{elhamdadi2022we}
\textsc{Elhamdadi H., Gaba A., Kim Y.-S., Xiong C.}:
\newblock How {D}o {W}e {M}easure {T}rust in {V}isual {D}ata {C}ommunication?
\newblock In \emph{2022 IEEE Evaluation and Beyond-Methodological Approaches for Visualization (BELIV)} (Piscataway, NJ, 2022), IEEE, pp.~85--92.

\bibitem[FEBL09]{faul2009statistical}
\textsc{Faul F., Erdfelder E., Buchner A., Lang A.-G.}:
\newblock Statistical {P}ower {A}nalyses using {G}* {P}ower 3.1: {T}ests for {C}orrelation and {R}egression {A}nalyses.
\newblock \emph{Behavior Research Methods 41}, 4 (2009), 1149--1160.

\bibitem[FELB07]{faul2007g}
\textsc{Faul F., Erdfelder E., Lang A.-G., Buchner A.}:
\newblock G* {P}ower 3: {A} {F}lexible {S}tatistical {P}ower {A}nalysis {P}rogram for the {S}ocial, {B}ehavioral, and {B}iomedical {S}ciences.
\newblock \emph{Behavior Research Methods 39}, 2 (2007), 175--191.

\bibitem[FGG90]{FurnhamGunterGreen1990}
\textsc{Furnham A., Gunter B., Green A.}:
\newblock Remembering {S}cience: {T}he {R}ecall of {F}actual {I}nformation as a {F}unction of the {P}resentation {M}ode.
\newblock \emph{Applied Cognitive Psychology 4}, 3 (1990), 203--212.

\bibitem[FPS{\etalchar{*}}21]{franconeri2021science}
\textsc{Franconeri S.~L., Padilla L.~M., Shah P., Zacks J.~M., Hullman J.}:
\newblock The {S}cience of {V}isual {D}ata {C}ommunication: {W}hat {W}orks.
\newblock \emph{Psychological Science in the Public Interest 22}, 3 (2021), 110--161.

\bibitem[FWM{\etalchar{*}}18]{fernandes2018uncertainty}
\textsc{Fernandes M., Walls L., Munson S., Hullman J., Kay M.}:
\newblock Uncertainty {D}isplays {U}sing {Q}uantile {D}otplots or {CDF}s {I}mprove {T}ransit {D}ecision-{M}aking.
\newblock In \emph{Proceedings of the 2018 CHI Conference on Human Factors in Computing Systems} (New York, NY, USA, 2018), Association for Computing Machinery, pp.~1--12.

\bibitem[GCBR{\etalchar{*}}21]{goodluck:2021}
\textsc{Goodluck~Constance T., Bajaj N., Rajwadi M., Maltby H., Wall J., Moniri M., Woodruff C., Laird T., Laird J., Glackin C., Cannings N.}:
\newblock Resolving {A}mbiguity in {H}edge {D}etection by {A}utomatic {G}eneration of {L}inguistic {R}ules.
\newblock In \emph{International Conference on Artificial Neural Networks} (2021), Springer-Verlag, p.~369–380.

\bibitem[GFVJ19]{GuyerEtAl2019}
\textsc{Guyer J.~J., Fabrigar L.~R., Vaughan-Johnston T.~I.}:
\newblock Speech {R}ate, {I}ntonation, and {P}itch: {I}nvestigating the {B}ias and {C}ue {E}ffects of {V}ocal {C}onfidence on {P}ersuasion.
\newblock \emph{Personality and Social Psychology Bulletin 45}, 3 (2019), 389--405.

\bibitem[Gil70]{Giles1970}
\textsc{Giles H.}:
\newblock Evaluative {R}eactions to {A}ccents.
\newblock \emph{Educational Review 22}, 3 (1970), 211--227.

\bibitem[GJS{\etalchar{*}}18]{greis:2018}
\textsc{Greis M., Joshi A., Singer K., Schmidt A., Machulla T.}:
\newblock Uncertainty {V}isualization {I}nfluences {H}ow {H}umans {A}ggregate {D}iscrepant {I}nformation.
\newblock In \emph{Proceedings of the 2018 CHI Conference on Human Factors in Computing Systems} (New York, NY, USA, 2018), CHI '18, Association for Computing Machinery, p.~1–12.
\newblock \href {https://doi.org/10.1145/3173574.3174079} {\path{doi:10.1145/3173574.3174079}}.

\bibitem[Goo23]{googlessml}
\textsc{Google}:
\newblock Speech {S}ynthesis {M}arkup {L}anguage ({SSML}) {R}eference, 2023.
\newblock Google Cloud Documentation.
\newblock URL: \url{https://cloud.google.com/text-to-speech/docs/ssml}.

\bibitem[GSWD17]{gortler:2017}
\textsc{G{\"o}rtler J., Schulz C., Weiskopf D., Deussen O.}:
\newblock Bubble {T}reemaps for {U}ncertainty {V}isualization.
\newblock \emph{IEEE Transactions on Visualization and Computer Graphics 24}, 1 (08 2017), 719--728.
\newblock \href {https://doi.org/10.1109/TVCG.2017.2743959} {\path{doi:10.1109/TVCG.2017.2743959}}.

\bibitem[HGH20]{hofman:2020}
\textsc{Hofman J.~M., Goldstein D.~G., Hullman J.}:
\newblock How {V}isualizing {I}nferential {U}ncertainty {C}an {M}islead {R}eaders {A}bout {T}reatment {E}ffects in {S}cientific {R}esults.
\newblock In \emph{Proceedings of the 2020 CHI Conference on Human Factors in Computing Systems} (New York, NY, USA, 2020), CHI '20, Association for Computing Machinery, p.~1–12.
\newblock \href {https://doi.org/10.1145/3313831.3376454} {\path{doi:10.1145/3313831.3376454}}.

\bibitem[Hla22]{stargazer2022package}
\textsc{Hlavac M.}:
\newblock \emph{{S}targazer: {W}ell-{F}ormatted {R}egression and {S}ummary {S}tatistics {T}ables}.
\newblock Social Policy Institute, Bratislava, Slovakia, 2022.
\newblock R package version 5.2.3.
\newblock URL: \url{https://CRAN.R-project.org/package=stargazer}.

\bibitem[HQC{\etalchar{*}}19]{hullmanpursuit:2019}
\textsc{Hullman J., Qiao X., Correll M., Kale A., Kay M.}:
\newblock In {P}ursuit of {E}rror: {A} {S}urvey of {U}ncertainty {V}isualization {E}valuation.
\newblock \emph{IEEE Transactions on Visualization and Computer Graphics 25}, 1 (jan 2019), 903–913.
\newblock \href {https://doi.org/10.1109/TVCG.2018.2864889} {\path{doi:10.1109/TVCG.2018.2864889}}.

\bibitem[HSSV22]{hoefer:2022}
\textsc{Hoefer M. J.~D., Schumacher B.~E., Szafir D.~A., Voida S.}:
\newblock Visualizing {U}ncertainty in {M}ulti-{S}ource {M}ental {H}ealth {D}ata.
\newblock In \emph{Extended Abstracts of the 2022 CHI Conference on Human Factors in Computing Systems} (New York, NY, USA, 2022), CHI EA '22, Association for Computing Machinery.
\newblock \href {https://doi.org/10.1145/3491101.3519844} {\path{doi:10.1145/3491101.3519844}}.

\bibitem[Hul16]{hullmanerrorprone:2016}
\textsc{Hullman J.}:
\newblock Why {E}valuating {U}ncertainty {V}isualization is {E}rror {P}rone.
\newblock In \emph{BELIV '16: Proceedings of the Sixth Workshop on Beyond Time and Errors on Novel Evaluation Methods for Visualization} (10 2016), Association for Computing Machinery, pp.~143--151.
\newblock \href {https://doi.org/10.1145/2993901.2993919} {\path{doi:10.1145/2993901.2993919}}.

\bibitem[IXM20]{islam-etal-2020-lexicon}
\textsc{Islam J., Xiao L., Mercer R.~E.}:
\newblock A {L}exicon-{B}ased {A}pproach for {D}etecting {H}edges in {I}nformal {T}ext.
\newblock In \emph{Proceedings of the 12th Language Resources and Evaluation Conference} (Marseille, France, 2020), European Language Resources Association, pp.~3109--3113.

\bibitem[JED{\etalchar{*}}20]{jena:2020}
\textsc{Jena A., Engelke U., Dwyer T., Raiamanickam V., Paris C.}:
\newblock Uncertainty {V}isualisation: {A}n {I}nteractive {V}isual {S}urvey.
\newblock In \emph{2020 IEEE Pacific Visualization Symposium (PacificVis)} (2020), IEEE, pp.~201--205.
\newblock \href {https://doi.org/10.1109/PacificVis48177.2020.1014} {\path{doi:10.1109/PacificVis48177.2020.1014}}.

\bibitem[Jen08]{Jensen2008}
\textsc{Jensen J.~D.}:
\newblock Scientific {U}ncertainty in {N}ews {C}overage of {C}ancer {R}esearch: {E}ffects of {H}edging on {S}cientists' and {J}ournalists' {C}redibility.
\newblock \emph{Human Communication Research 34}, 3 (2008), 347--369.

\bibitem[JL12]{joslyn2012uncertainty}
\textsc{Joslyn S.~L., LeClerc J.~E.}:
\newblock Uncertainty {F}orecasts {I}mprove {W}eather-{R}elated {D}ecisions and {A}ttenuate the {E}ffects of {F}orecast {E}rror.
\newblock \emph{Journal of Experimental Psychology: Applied 18}, 1 (2012), 126--140.

\bibitem[JP15]{JiangPell2015}
\textsc{Jiang X., Pell M.~D.}:
\newblock On {H}ow the {B}rain {D}ecodes {V}ocal {C}ues {A}bout {S}peaker {C}onfidence.
\newblock \emph{Cortex 66} (2015), 9--34.

\bibitem[JP17]{JiangPell2017}
\textsc{Jiang X., Pell M.~D.}:
\newblock The {S}ound of {C}onfidence and {D}oubt.
\newblock \emph{Speech Communication 88} (2017), 106--126.

\bibitem[JS03]{johnson:2003}
\textsc{Johnson C., Sanderson A.}:
\newblock A {N}ext {S}tep: {V}isualizing {E}rrors and {U}ncertainty.
\newblock \emph{Computer Graphics and Applications, IEEE 23} (09 2003), 6--10.
\newblock \href {https://doi.org/10.1109/MCG.2003.1231171} {\path{doi:10.1109/MCG.2003.1231171}}.

\bibitem[Kay23]{ggdist}
\textsc{Kay M.}:
\newblock \emph{{ggdist}: {V}isualizations of {D}istributions and {U}ncertainty}.
\newblock Northwestern University, 2023.
\newblock R package version 3.3.0.
\newblock URL: \url{https://mjskay.github.io/ggdist/}, \href {https://doi.org/10.5281/zenodo.3879620} {\path{doi:10.5281/zenodo.3879620}}.

\bibitem[KCG15]{Koo2015GeovisualizationOA}
\textsc{Koo H., Chun Y., Griffith D.~A.}:
\newblock Geovisualization of {A}ttribute {U}ncertainty.
\newblock \emph{Journal of Visual Languages \& Computing 44} (2015), 89--96.

\bibitem[KKGMH20]{kim2020bayesian}
\textsc{Kim Y.-S., Kayongo P., Grunde-McLaughlin M., Hullman J.}:
\newblock Bayesian-assisted {I}nference from {V}isualized {D}ata.
\newblock \emph{IEEE Transactions on Visualization and Computer Graphics 27}, 2 (2020), 989--999.

\bibitem[KKH21]{kalevisualreasoning:2021}
\textsc{Kale A., Kay M., Hullman J.}:
\newblock Visual {R}easoning {S}trategies for {E}ffect {S}ize {J}udgments and {D}ecisions.
\newblock \emph{IEEE Transactions on Visualization \& Computer Graphics 27}, 02 (Feb 2021), 272--282.
\newblock \href {https://doi.org/10.1109/TVCG.2020.3030335} {\path{doi:10.1109/TVCG.2020.3030335}}.

\bibitem[KKHM16]{kay2016ish}
\textsc{Kay M., Kola T., Hullman J.~R., Munson S.~A.}:
\newblock When ({I}sh) is {M}y {B}us? {U}ser-{C}entered {V}isualizations of {U}ncertainty in {E}veryday, {M}obile {P}redictive {S}ystems.
\newblock In \emph{Proceedings of the 2016 CHI Conference on Human Factors in Computing Systems} (New York, NY, USA, 2016), Association for Computing Machinery, pp.~5092--5103.

\bibitem[KKS{\etalchar{*}}75]{KintschEtAl1975}
\textsc{Kintsch W., Kozminsky E., Streby W.~J., McKoon G., Keenan J.~M.}:
\newblock Comprehension and {R}ecall of {T}ext as a {F}unction of {C}ontent {V}ariables.
\newblock \emph{Journal of Verbal Learning and Verbal Behavior 14}, 2 (1975), 196--214.

\bibitem[KLSG22]{kirkland2022s}
\textsc{Kirkland A., Lameris H., Sz{\'e}kely E., Gustafson J.}:
\newblock Where’s the {U}h, {H}esitation? {T}he {I}nterplay {B}etween {F}illed {P}ause {L}ocation, {S}peech {R}ate and {F}undamental {F}requency in {P}erception of {C}onfidence.
\newblock In \emph{Proceedings of Interspeech} (Incheon, Korea, 2022), Interspeech, pp.~4990--4994.

\bibitem[KN22]{moments}
\textsc{Komsta L., Novomestky F.}:
\newblock \emph{Moments, {C}umulants, {S}kewness, {K}urtosis and {R}elated {T}ests}.
\newblock CRAN, 2022.
\newblock R package version 0.14.1.
\newblock URL: \url{https://cran.r-project.org/web/packages/moments/moments.pdf}.

\bibitem[KRF20]{Korporaal2020EffectsOU}
\textsc{Korporaal M., Ruginski I.~T., Fabrikant S.~I.}:
\newblock Effects of {U}ncertainty {V}isualization on {M}ap-{B}ased {D}ecision {M}aking {U}nder {T}ime {P}ressure.
\newblock \emph{Frontiers of Computer Science 2} (2020), Article 32.

\bibitem[KS11]{KatzSelkirk2011}
\textsc{Katz J., Selkirk E.}:
\newblock Contrastive {F}ocus vs. {D}iscourse-new: {E}vidence {F}rom {P}honetic {P}rominence in {English}.
\newblock \emph{Language 87}, 4 (2011), 771--816.

\bibitem[KWKH19]{kim2019bayesian}
\textsc{Kim Y.-S., Walls L.~A., Krafft P., Hullman J.}:
\newblock A {B}ayesian {C}ognition {A}pproach to {I}mproving {D}ata {V}isualization.
\newblock In \emph{Proceedings of the 2019 CHI Conference on Human Factors in Computing Systems} (2019), pp.~1--14.

\bibitem[Lak73]{lakoff:1973}
\textsc{Lakoff G.}:
\newblock Hedges: {A} {S}tudy in {M}eaning {C}riteria and the {L}ogic of {F}uzzy {C}oncepts.
\newblock \emph{Journal of Philosophical Logic 2}, 4 (1973), 458--508.
\newblock \href {https://doi.org/10.1007/BF00262952} {\path{doi:10.1007/BF00262952}}.

\bibitem[LBR{\etalchar{*}}16]{liu:2016}
\textsc{Liu L., Boone A., Ruginski I., Padilla L., Hegarty M., Creem-Regehr S., Thompson W., Yuksel C., House D.}:
\newblock Uncertainty {V}isualization by {R}epresentative {S}ampling from {P}rediction {E}nsembles.
\newblock \emph{IEEE transactions on visualization and computer graphics PP} (09 2016).
\newblock \href {https://doi.org/10.1109/TVCG.2016.2607204} {\path{doi:10.1109/TVCG.2016.2607204}}.

\bibitem[LM21]{leffrang:2021}
\textsc{Leffrang D., M\"{u}ller O.}:
\newblock Should {I} {F}ollow this {M}odel? {T}he {E}ffect of {U}ncertainty {V}isualization on the {A}cceptance of {T}ime {S}eries {F}orecasts.
\newblock In \emph{2021 IEEE Workshop on TRust and EXpertise in Visual Analytics (TREX)} (2021), IEEE, pp.~20--26.
\newblock \href {https://doi.org/10.1109/TREX53765.2021.00009} {\path{doi:10.1109/TREX53765.2021.00009}}.

\bibitem[LMT71]{LondonEtAl1971}
\textsc{London H., McSeveney D., Tropper R.}:
\newblock Confidence, {O}verconfidence and {P}ersuasion.
\newblock \emph{Human Relations 24}, 5 (1971), 359--369.

\bibitem[LP84]{LibermanPierrehumbert1984}
\textsc{Liberman M., Pierrehumbert J.}:
\newblock Intonational {I}nvariance {U}nder {C}hanges in {P}itch {R}ange and {L}ength.
\newblock In \emph{Language Sound Structure. Studies in phonology presented to {Morris Halle}}, Aronoff M., Oehrle R.~T., (Eds.). MIT Press, Cambridge, MA, 1984, pp.~157--233.

\bibitem[LW10]{LasarcykWollermann2010}
\textsc{Lasarcyk E., Wollermann C.}:
\newblock Do {P}rosodic {C}ues {I}nfluence {U}ncertainty {P}erception in {A}rticulatory {S}peech {S}ynthesis?
\newblock In \emph{Seventh ISCA Workshop on Speech Synthesis} (Kyoto, Japan, 2010), nternational Symposium on Computer Architecture (ISCA), pp.~230--235.

\bibitem[MDL08]{Morss2008CommunicatingUI}
\textsc{Morss R., Demuth J.~L., Lazo J.~K.}:
\newblock Communicating {U}ncertainty in {W}eather {F}orecasts: {A} {S}urvey of the {U.S.} {P}ublic.
\newblock \emph{Weather and Forecasting 23} (2008), 974--991.
\newblock \href {https://doi.org/10.1175/2008WAF2007088.1} {\path{doi:10.1175/2008WAF2007088.1}}.

\bibitem[MH90]{Morgan1990UncertaintyAG}
\textsc{Morgan M.~G., Henrion M.}:
\newblock \emph{Uncertainty: {A} {G}uide to {D}ealing with {U}ncertainty in {Q}uantitative {R}isk and {P}olicy {A}nalysis}.
\newblock Cambridge University Press, Cambridge, 1990.
\newblock URL: \url{https://api.semanticscholar.org/CorpusID:57858998}.

\bibitem[MLB{\etalchar{*}}20]{mulder2020designing}
\textsc{Mulder K.~J., Lickiss M., Black A., Charlton-Perez A.~J., McCloy R., Young J.~S.}:
\newblock Designing {E}nvironmental {U}ncertainty {I}nformation for {E}xperts and {N}on-experts: {D}oes {D}ata {P}resentation {A}ffect {U}sers’ {D}ecisions and {I}nterpretations?
\newblock \emph{Meteorological Applications 27}, 1 (2020), Article e1821.

\bibitem[MMBV76]{MillerEtAl1976}
\textsc{Miller N., Maruyama G., Beaber R.~J., Valone K.}:
\newblock Speed of {S}peech and {P}ersuasion.
\newblock \emph{Journal of Personality and Social Psychology 34}, 4 (1976), 615--624.

\bibitem[MR77]{MulacRudd1977}
\textsc{Mulac A., Rudd M.~J.}:
\newblock Effects of {S}elected {American} {R}egional {D}ialects upon {R}egional {A}udience {M}embers.
\newblock \emph{Communications Monographs 44}, 3 (1977), 185--195.

\bibitem[MRH{\etalchar{*}}05]{MacEachren:2005}
\textsc{MacEachren A., Robinson A., Hopper S., Gardner S., Murray R., Gahegan M., Hetzler E.}:
\newblock Visualizing {G}eospatial {I}nformation {U}ncertainty: {W}hat {W}e {K}now and {W}hat {W}e {N}eed to {K}now.
\newblock \emph{Cartography and Geographic Information Science 32} (07 2005), 139--160.
\newblock \href {https://doi.org/10.1559/1523040054738936} {\path{doi:10.1559/1523040054738936}}.

\bibitem[MRO{\etalchar{*}}12]{maceachren2012visual}
\textsc{MacEachren A.~M., Roth R.~E., O'Brien J., Li B., Swingley D., Gahegan M.}:
\newblock Visual {S}emiotics \& {U}ncertainty {V}isualization: {A}n {E}mpirical {S}tudy.
\newblock \emph{IEEE Transactions on Visualization and Computer Graphics 18}, 12 (2012), 2496--2505.

\bibitem[MW69]{MehrabianWilliams1969}
\textsc{Mehrabian A., Williams M.}:
\newblock Nonverbal {C}oncomitants of {P}erceived and {I}ntended {P}ersuasiveness.
\newblock \emph{Journal of Personality and Social psychology 13}, 1 (1969), 37--58.

\bibitem[NCA{\etalchar{*}}19]{nagel:2019}
\textsc{Nagel F., Castiglia G., Ademaj G., Buchmller J., Schlegel U., Keim D.~A.}:
\newblock {cpmViz}: {A} {W}eb-{B}ased {V}isualization {T}ool for {U}ncertain {S}patiotemporal {D}ata.
\newblock In \emph{2019 IEEE Conference on Visual Analytics Science and Technology (VAST)} (2019), IEEE, pp.~140--141.
\newblock \href {https://doi.org/10.1109/VAST47406.2019.8986941} {\path{doi:10.1109/VAST47406.2019.8986941}}.

\bibitem[NGJ09]{nadav2009uncertainty}
\textsc{Nadav-Greenberg L., Joslyn S.~L.}:
\newblock Uncertainty {F}orecasts {I}mprove {D}ecision {M}aking {A}mong {N}onexperts.
\newblock \emph{Journal of Cognitive Engineering and Decision Making 3}, 3 (2009), 209--227.

\bibitem[OJGW19]{okan2019using}
\textsc{Okan Y., Janssen E., Galesic M., Waters E.~A.}:
\newblock Using the {S}hort {G}raph {L}iteracy {S}cale to {P}redict {P}recursors of {H}ealth {B}ehavior {C}hange.
\newblock \emph{Medical Decision Making 39}, 3 (2019), 183--195.

\bibitem[Pan01]{Pang2001VisualizingUI}
\textsc{Pang A.~T.}:
\newblock Visualizing {U}ncertainty in {G}eospatial {D}ata.
\newblock In \emph{Proceedings of the Workshop on the Intersections between Geospatial Information and Information Technology} (2001).
\newblock URL: \url{https://api.semanticscholar.org/CorpusID:16285011}.

\bibitem[PBS11]{PonBarrySchieber2011}
\textsc{Pon-Barry H., Shieber S.~M.}:
\newblock Recognizing {U}ncertainty in {S}peech.
\newblock \emph{EURASIP Journal on Advances in Signal Processing 2011} (2011), Article 251753.

\bibitem[PCH21]{PADILLA2021275}
\textsc{Padilla L., Castro S.~C., Hosseinpour H.}:
\newblock Chapter {S}even - {A} {R}eview of {U}ncertainty {V}isualization {E}rrors: {W}orking {M}emory as an {E}xplanatory {T}heory.
\newblock In \emph{The Psychology of Learning and Motivation}, Federmeier K.~D., (Ed.), vol.~74 of \emph{Psychology of Learning and Motivation}. Academic Press, Cambridge, MA, 2021, pp.~275--315.
\newblock URL: \url{https://www.sciencedirect.com/science/article/pii/S0079742121000074}, \href {https://doi.org/10.1016/bs.plm.2021.03.001} {\path{doi:10.1016/bs.plm.2021.03.001}}.

\bibitem[PCRHS18]{padilla2018decision}
\textsc{Padilla L.~M., Creem-Regehr S.~H., Hegarty M., Stefanucci J.~K.}:
\newblock Decision {M}aking with {V}isualizations: {A} {C}ognitive {F}ramework {A}cross {D}isciplines.
\newblock \emph{Cognitive Research: Principles and Implications 3}, 1 (2018), Article 29.

\bibitem[PHF{\etalchar{*}}22]{padillacovid19:2022}
\textsc{Padilla L., Hosseinpour H., Fygenson R., Howell J., Chunara R., Bertini E.}:
\newblock Impact of {COVID-19} {F}orecast {V}isualizations on {P}andemic {R}isk {P}erceptions.
\newblock \emph{Scientific Reports 12}, 1 (Feb. 2022), Article 2014.
\newblock Funding Information: The National Science Foundation supported this work with Grant 2028374. Publisher Copyright: {\textcopyright} 2022, The Author(s).
\newblock \href {https://doi.org/10.1038/s41598-022-05353-1} {\path{doi:10.1038/s41598-022-05353-1}}.

\bibitem[PKH21]{padillabook:2021}
\textsc{Padilla L., Kay M., Hullman J.}:
\newblock Uncertainty {V}isualization.
\newblock In \emph{Wiley StatsRef: Statistics Reference Online}. Wiley, 02 2021, pp.~1--18.
\newblock \href {https://doi.org/10.1002/9781118445112.stat08296} {\path{doi:10.1002/9781118445112.stat08296}}.

\bibitem[PMCO23]{pandey2023you}
\textsc{Pandey S., McKinley O.~G., Crouser R.~J., Ottley A.}:
\newblock Do {Y}ou {T}rust {W}hat {Y}ou {S}ee? {T}oward {A} {M}ultidimensional {M}easure of {T}rust in {V}isualization.
\newblock \emph{IEEE Transactions on Visualization and Computer Graphics} (2023).

\bibitem[PPKH21]{padilla2021uncertain}
\textsc{Padilla L.~M., Powell M., Kay M., Hullman J.}:
\newblock Uncertain about {U}ncertainty: {H}ow {Q}ualitative {E}xpressions of {F}orecaster {C}onfidence {I}mpact {D}ecision-making with {U}ncertainty {V}isualizations.
\newblock \emph{Frontiers in Psychology 11} (2021), 579267.

\bibitem[PRCR17]{padillaensemble:2017}
\textsc{Padilla L., Ruginski I., Creem-Regehr S.}:
\newblock Effects of {E}nsemble and {S}ummary {D}isplays on {I}nterpretations of {G}eospatial {U}ncertainty {D}ata.
\newblock \emph{Cognitive Research: Principles and Implications 2} (10 2017).
\newblock \href {https://doi.org/10.1186/s41235-017-0076-1} {\path{doi:10.1186/s41235-017-0076-1}}.

\bibitem[PS18]{palan2018prolific}
\textsc{Palan S., Schitter C.}:
\newblock Prolific.ac — {A} {S}ubject {P}ool for {O}nline {E}xperiments.
\newblock \emph{Journal of Behavioral and Experimental Finance 17} (2018), 22--27.

\bibitem[{R C}21]{RComputing}
\textsc{{R Core Team}}:
\newblock \emph{R: {A} {L}anguage and {E}nvironment for {S}tatistical {C}omputing}.
\newblock R Foundation for Statistical Computing, Vienna, Austria, 2021.
\newblock \url{https://www.R-project.org/}.

\bibitem[RLK06]{rubinbook}
\textsc{Rubin V., Liddy E., Kando N.}:
\newblock Certainty {I}dentification in {T}exts: {C}ategorization {M}odel and {M}anual {T}agging {R}esults.
\newblock In \emph{Computing Attitude and Affect in Text: Theory and Applications. The Information Retrieval Series}, Shanahan J.~G., Qu Y., Wiebe J., (Eds.), vol.~20. Springer, Dordrecht, 01 2006, pp.~61--76.
\newblock \href {https://doi.org/10.1007/1-4020-4102-0_7} {\path{doi:10.1007/1-4020-4102-0_7}}.

\bibitem[RPHL14]{RISTOVSKI201460}
\textsc{Ristovski G., Preusser T., Hahn H.~K., Linsen L.}:
\newblock Uncertainty in {M}edical {V}isualization: {T}owards a {T}axonomy.
\newblock \emph{Computers \& Graphics 39} (2014), 60--73.
\newblock URL: \url{https://www.sciencedirect.com/science/article/pii/S0097849313001568}, \href {https://doi.org/10.1016/j.cag.2013.10.015} {\path{doi:10.1016/j.cag.2013.10.015}}.

\bibitem[SAC98]{SparksAreniCox1998}
\textsc{Sparks J.~R., Areni C.~S., Cox K.~C.}:
\newblock An {I}nvestigation of the {E}ffects of {L}anguage {S}tyle and {C}ommunication {M}odality on {P}ersuasion.
\newblock \emph{Communications Monographs 65}, 2 (1998), 108--125.

\bibitem[SAJ21]{Sane2021VisualizationOU}
\textsc{Sane S., Athawale T.~M., Johnson C.~R.}:
\newblock Visualization of {U}ncertain {M}ultivariate {D}ata via {F}eature {C}onfidence {L}evel-{S}ets.
\newblock In \emph{Eurographics Conference on Visualization} (2021), The Eurographics Association, pp.~43--47.

\bibitem[SC93]{SmithClark1993}
\textsc{Smith V.~L., Clark H.~H.}:
\newblock On the {C}ourse of {A}nswering {Q}uestions.
\newblock \emph{Journal of Memory and Language 32} (1993), 25--38.

\bibitem[SC22]{setlur2022functional}
\textsc{Setlur V., Cogley B.}:
\newblock \emph{Functional Aesthetics for Data Visualization}.
\newblock John Wiley \& Sons, 2022.

\bibitem[Sch97]{schriver:1997}
\textsc{Schriver K.~A.}:
\newblock \emph{Dynamics in {D}ocument {D}esign: {C}reating {T}ext for {R}eaders}.
\newblock John Wiley \& Sons, Inc., USA, 1997.

\bibitem[SJ13]{savelli2013advantages}
\textsc{Savelli S., Joslyn S.}:
\newblock The {A}dvantages of {P}redictive {I}nterval {F}orecasts for {N}on-expert {U}sers and the {I}mpact of {V}isualizations.
\newblock \emph{Applied Cognitive Psychology 27}, 4 (2013), 527--541.

\bibitem[SJW00]{smith2000users}
\textsc{Smith-Jackson T.~L., Wogalter M.~S.}:
\newblock Users' {H}azard {P}erceptions of {W}arning {C}omponents: {A}n {E}xamination of {C}olors and {S}ymbols.
\newblock In \emph{Proceedings of the Human Factors and Ergonomics Society Annual Meeting} (2000), vol.~44, SAGE Publications, pp.~6--55.

\bibitem[SK05]{SwertsKrahmer2005}
\textsc{Swerts M., Krahmer E.}:
\newblock Audiovisual {P}rosody and {F}eeling of {K}nowing.
\newblock \emph{Journal of Memory and Language 53} (2005), 81--94.

\bibitem[SLSR08]{skeels:2008}
\textsc{Skeels M., Lee B., Smith G., Robertson G.}:
\newblock Revealing {U}ncertainty for {I}nformation {V}isualization.
\newblock In \emph{AVI '08: Proceedings of the working conference on advanced visual interfaces} (05 2008), vol.~9, pp.~376--379.
\newblock \href {https://doi.org/10.1057/ivs.2009.1} {\path{doi:10.1057/ivs.2009.1}}.

\bibitem[SLW73]{SchererEtAl1973}
\textsc{Scherer K.~R., London H., Wolf J.~J.}:
\newblock The {V}oice of {C}onfidence: {P}aralinguistic {C}ues and {A}udience {E}valuation.
\newblock \emph{Journal of Research in Personality 7} (1973), 31--44.

\bibitem[Sor97]{Sorensen1997}
\textsc{Sorensen R.}:
\newblock Vagueness.
\newblock In \emph{Stanford Encyclopedia of Philosophy}, Zalta E.~N., (Ed.). Stanford University, 1997.

\bibitem[SPS11]{Spiegelhalter:20111}
\textsc{Spiegelhalter D., Pearson M., Short I.}:
\newblock Visualizing {U}ncertainty {A}bout the {F}uture.
\newblock \emph{Science 333}, 6048 (2011), 1393--1400.
\newblock \href {https://doi.org/10.1126/science.1191181} {\path{doi:10.1126/science.1191181}}.

\bibitem[SRG85]{StewartRyanGiles1985}
\textsc{Stewart M.~A., Ryan E.~B., Giles H.}:
\newblock Accent and {S}ocial {C}lass {E}ffects on {S}tatus and {S}olidarity {E}valuations.
\newblock \emph{Personality and Social Psychology Bulletin 11}, 1 (1985), 98--105.

\bibitem[SSK{\etalchar{*}}16]{sacha:2016}
\textsc{Sacha D., Senaratne H., Kwon B.~C., Ellis G., Keim D.~A.}:
\newblock The {R}ole of {U}ncertainty, {A}wareness, and {T}rust in {V}isual {A}nalytics.
\newblock \emph{IEEE Transactions on Visualization and Computer Graphics 22}, 1 (2016), 240--249.
\newblock \href {https://doi.org/10.1109/TVCG.2015.2467591} {\path{doi:10.1109/TVCG.2015.2467591}}.

\bibitem[SSR21]{SteijaertSchaapRiet2021}
\textsc{Steijaert M.~J., Schaap G., Riet J.~V.}:
\newblock Two-sided {S}cience: {C}ommunicating {S}cientific {U}ncertainty {I}ncreases {T}rust in {S}cientists and {D}onation {I}ntention by {D}ecreasing {A}ttribution of {C}ommunicator {B}ias.
\newblock \emph{Communications 46}, 2 (2021), 297--316.

\bibitem[SVF{\etalchar{*}}12]{SzarvasEtAl2012}
\textsc{Szarvas G., Vincze V., Farkas R., M{\'o}ra G., Gurevych I.}:
\newblock Cross-genre and {C}ross-domain {D}etection of {S}emantic {U}ncertainty.
\newblock \emph{Computational Linguistics 38}, 2 (2012), 335--367.

\bibitem[SW95]{Sperber1995-SPER}
\textsc{Sperber D., Wilson D.}:
\newblock \emph{Relevance: {C}ommunication and {C}ognition}.
\newblock Blackwell, Oxford, 1986/1995.

\bibitem[THM{\etalchar{*}}05]{Thomson2005ATF}
\textsc{Thomson J.~R., Hetzler E.~G., MacEachren A.~M., Gahegan M., Pavel M.}:
\newblock A {T}ypology for {V}isualizing {U}ncertainty.
\newblock In \emph{Proceedings of the Society of Photo-Optical Instrumentation Engineers (SPIE) 5669} (2005), Erbacher R.~F., Roberts J.~C., Grohn M.~T., Borner K., (Eds.), SPIE, pp.~146--157.

\bibitem[TKH08]{tsai2008effects}
\textsc{Tsai C.~I., Klayman J., Hastie R.}:
\newblock Effects of {A}mount of {I}nformation on {J}udgment {A}ccuracy and {C}onfidence.
\newblock \emph{Organizational Behavior and Human Decision Processes 107}, 2 (2008), 97--105.

\bibitem[Tou03]{toulmin_2003}
\textsc{Toulmin S.~E.}:
\newblock \emph{The {U}ses of {A}rgument}, second~ed.
\newblock Cambridge University Press, Cambridge, 2003.
\newblock \href {https://doi.org/10.1017/CBO9780511840005} {\path{doi:10.1017/CBO9780511840005}}.

\bibitem[TT14]{tak2014color}
\textsc{Tak S., Toet A.}:
\newblock Color and {U}ncertainty: {I}t is {N}ot {A}lways {B}lack and {W}hite.
\newblock In \emph{Eurographics Conference on Visualization (EuroVis)} (2014), The Eurographics Association, pp.~55--59.

\bibitem[VKKG17]{vosough:2017}
\textsc{Vosough Z., Kammer D., Keck M., Groh R.}:
\newblock Visualizing {U}ncertainty in {F}low {D}iagrams: {A} {C}ase {S}tudy in {P}roduct {C}osting.
\newblock In \emph{Proceedings of the 10th International Symposium on Visual Information Communication and Interaction} (New York, NY, USA, 2017), VINCI '17, Association for Computing Machinery, pp.~1--8.
\newblock \href {https://doi.org/10.1145/3105971.3105972} {\path{doi:10.1145/3105971.3105972}}.

\bibitem[VZB20]{VanZantBerger2020}
\textsc{Van~Zant A.~B., Berger J.}:
\newblock How the {V}oice {P}ersuades.
\newblock \emph{Journal of Personality and Social Psychology 118}, 4 (2020), 661--682.

\bibitem[WII{\etalchar{*}}12]{wood2012sketchy}
\textsc{Wood J., Isenberg P., Isenberg T., Dykes J., Boukhelifa N., Slingsby A.}:
\newblock Sketchy {R}endering for {I}nformation {V}isualization.
\newblock \emph{IEEE Transactions on Visualization and Computer Graphics 18}, 12 (2012), 2749--2758.

\bibitem[WLWC23]{witt:2023}
\textsc{Witt J.~K., Labe Z.~M., Warden A.~C., Clegg B.~A.}:
\newblock Visualizing {U}ncertainty in {H}urricane {F}orecasts with {A}nimated {R}isk {T}rajectories.
\newblock \emph{Weather, Climate, and Society 15}, 2 (2023), 407 -- 424.
\newblock \href {https://doi.org/10.1175/WCAS-D-21-0173.1} {\path{doi:10.1175/WCAS-D-21-0173.1}}.

\bibitem[XPGF19]{xiong2019examining}
\textsc{Xiong C., Padilla L., Grayson K., Franconeri S.}:
\newblock Examining the {C}omponents of {T}rust in {M}ap-based {V}isualizations.
\newblock In \emph{1st EuroVis Workshop on Trustworthy Visualization, TrustVis 2019} (Genova, Italy, 2019), The Eurographics Association, pp.~19--23.

\end{thebibliography}


%
%

\end{document}